\documentclass[aps, prd, amsmath, floats,floatfix, superscriptaddress, nofootinbib, showkeys]{revtex4}

\usepackage{amssymb}
\usepackage{amsmath}
\usepackage{verbatim}
\usepackage{mathrsfs}
\usepackage{amsfonts}
\usepackage{amsfonts} 
\usepackage{latexsym}
\usepackage{epsfig}
\usepackage{color}
\usepackage{graphicx,subfigure}
\usepackage{units}
\usepackage{envmath}
\usepackage{natbib}
\usepackage{tikz}
\usepackage{soul}
\usepackage{ulem}
\usepackage{bigints}
\usepackage{makecell}

\begin{document}
	
	
	\definecolor{orange}{rgb}{0.9,0.45,0}
	
	\newcommand{\re}{\mbox{Re}}
	\newcommand{\im}{\mbox{Im}}

	\def\CovDev{D}
	\def\Res{{\mathcal R}}
	\def\Gammaflat{\hat \Gamma}
	\def\metricflat{\hat \gamma}
	\def\Dflat{\hat {\mathcal D}}
	\def\part_n{\partial_\perp}
	
	\def\Lie{\mathcal{L}}
	\def\A{\mathcal{X}}
	\def\Aphi{\A_{\phi}}
	\def\hAphi{\hat{\A}_{\phi}}
	\def\E{\mathcal{E}}
	\def\Ham{\mathcal{H}}
	\def\M{\mathcal{M}}
	\def\R{\mathcal{R}}
	\def\p{\partial}
	
	\def\hg{\hat{\gamma}}
	\def\hA{\hat{A}}
	\def\hD{\hat{D}}
	\def\hE{\hat{E}}
	\def\hR{\hat{R}}
	\def\hcA{\hat{\mathcal{A}}}
	\def\hDelt{\hat{\triangle}}
	
	\def\na{\nabla}
	\def\dif{{\rm{d}}}
	\def\non{\nonumber}
	\newcommand{\erf}{\textrm{erf}}
	
	\renewcommand{\t}{\times}
	
	\long\def\symbolfootnote[#1]#2{\begingroup%
		\def\thefootnote{\fnsymbol{footnote}}\footnote[#1]{#2}\endgroup}

 \newcommand{\EDV}[1]{\textcolor{green}{[{\bf EDV}: #1]}} 

\title{A new binning method to choose a standard set of Quasars}


\author{M.G. Dainotti}
\email{maria.dainotti@nao.ac.jp}
\affiliation{National Astronomical Observatory of Japan,\\ 2 Chome-21-1 Osawa, Mitaka, Tokyo 181-8588, Japan}
\affiliation{The Graduate University for Advanced Studies, SOKENDAI,\\ Shonankokusaimura, Hayama, Miura District, Kanagawa 240-0193, Japan}
\affiliation{Space Science Institute,\\ 4765 Walnut St, Suite B, 80301 Boulder, CO, USA}
\affiliation{Department of Physics and Astrophysics, University of Las Vegas, \\ NV 89154, USA}
\author{A. {\L}. Lenart}
\email{aleksander.lenart@student.uj.edu.pl}
\affiliation{Astronomical Observatory of Jagiellonian University in Krak{\'o}w,\\ ul. Orla 171, 30-244 Krak{\'o}w, Poland}
\author{M. Ghodsi Yengejeh}
\email{mina.ghodsi@csfk.org}
\affiliation{PDAT Laboratory, Department of Physics, K.N. Toosi University of Technology,\\ Tehran, Iran}
\affiliation{Konkoly Observatory, HUN-REN Research Centre of Astronomy and Earth Sciences (CSFK), MTA Centre of Excellence, Budapest, Konkoly Thege Mikl{\'o}s {\'u}t 15-17. H-1121 Hungary}
\affiliation{MTA-CSFK Lend\"ulet Large-scale Structure Research Group,  H-1121 Budapest, Konkoly Thege Mikl\'os \'ut 15-17, Hungary}
\author{S. Chakraborty}
\email{s06schak@uni-bonn.de}
\affiliation{Department of Physics, Universit\"at Bonn, Nussallee 12,
D-53115 Bonn, Germany}
\author{Nissim Fraija}
\email{nifraija@astro.unam.mx}
\affiliation{Instituto de Astronom{\'i}a, Universidad Nacional Aut{\'o}noma de M{\'e}xico, Circuito Exterior, C.U., A. Postal 70-264, 04510 M{\'e}xico City, M{\'e}xico}
\author{E. Di Valentino}
\email{e.divalentino@sheffield.ac.uk}
\affiliation{School of Mathematics and Statistics, University of Sheffield, Hounsfield Road, Sheffield S3 7RH, United Kingdom}
\author{G. Montani}
\email{giovanni.montani@enea.it}
\affiliation{ENEA, Fusion and Nuclear Safety Department, C. R. Frascati, Via E. Fermi 45, 00044 Frascati (RM), Italy}
\affiliation{Department of Physics, "La Sapienza" University of Rome, P.le Aldo Moro 5, 00185 Rome, Italy}

	
\date{\today}
	
	
\begin{abstract} 
Although the Lambda Cold Dark Matter model is the most accredited cosmological model, information at intermediate redshifts (z) between type Ia Supernovae (z = 2.26) and the Cosmic Microwave Background (z = 1100) is crucial to validate this model further. Here, we present a detailed and reliable methodology for binning the quasars (QSO) data that allows the identification of a golden sample of QSOs to be used as standard candles. This procedure has the advantage of being very general. Thus, it can be applied to any astrophysical sources at cosmological distances.
This methodology allows us to avoid the circularity problem since it involves a flux-flux relation and includes the analysis of removing selection biases and the redshift evolution. With this method, we have discovered a sample of 1253 quasars up to z = 7.54 with reduced intrinsic dispersion of the relation between Ultraviolet and X-ray fluxes, with $\delta_{int} = 0.096\pm 0.003$ (56\% less than the original sample where $\delta_{int} =0.22$). Once the luminosities are corrected for selection biases and redshift evolution, this `gold' sample allows us to determine the matter density parameter to be $\Omega_M=0.240 \pm 0.064$. This value is aligned with the results of the $\Lambda CDM$ model obtained with SNe Ia. 
\end{abstract}
	\maketitle
	\vspace{0.8cm}

\maketitle

\section{Introduction}
Recently, the improved precision of the measurements of cosmological parameters has uncovered significant inconsistencies in the widely accepted cosmological model known as the flat $\Lambda$ Cold Dark Matter ($\Lambda$CDM) model. This model describes the Universe by incorporating a CDM and dark energy components, with the dark energy represented as a cosmological constant ($\Lambda$), as required by the current accelerated expansion of the Universe~\cite{riess1998, perlmutter1999}. 
The flat $\Lambda$CDM model successfully fits
observations such as the cosmic microwave background (CMB) radiation~\cite{Planck:2018vyg}, the baryonic acoustic oscillations (BAO)~\cite{eboss2021}, and the acceleration of the expansion of the Universe proved by type Ia supernovae (SNe Ia). Besides that, it presents very well-known theoretical shortcomings. This is the case of the cosmological constant problem~\cite{1989RvMP...61....1W}, which is the tension between the expected and the observed values of $\Omega_{\Lambda}$, the nature of dark energy and its origin, and the fine-tuning (or coincidence) problem, which derives from the fact that the current values of the matter density ($\Omega_{M}$) and the dark energy density ($\Omega_{\Lambda}$) are of the same order, whereas this is not expected due to their different evolution in time. 
In addition to these issues, recent measurements have highlighted the so-called Hubble constant ($H_{0}$) tension. This is the discrepancy between the value of $H_{0}$ measured locally from SNe Ia and Cepheids, which is $H_{0} = 73.04 \pm 1.04\, \rm km \, s^{-1}\, Mpc^{-1}$~\cite{2022ApJ...934L...7R}, and the value of $H_{0}$ predicted from the Planck data on the CMB within a flat $\Lambda$CDM model, $H_{0} = 67.4 \pm 0.5\, \rm km\,s^{-1}\, Mpc^{-1}$~\cite{Planck:2018vyg}. The difference between these two measurements ranges between 4.4 to 6 $\sigma$, according to the samples investigated~\cite{Riess2019ApJ...876...85R, 2020PhRvR...2a3028C, 2020MNRAS.498.1420W,Abdalla:2022yfr,DiValentino:2021izs,DiValentino:2020zio,Verde:2019ivm,Riess:2019qba,DiValentino:2022fjm,DiValentino:2020vnx}. However, the maximum redshift reached by SNe Ia observations is z = 2.26~\cite{Rodney}, while the CMB radiation is observed at z = 1100. 
Thus, it is crucial to probe the Universe in the intermediate epochs between these two redshifts to shed light on this tension, hence to confirm, alleviate, or even solve it. To this end, other probes rather than SNe Ia and CMB have already been investigated. These analyses have provided even more complicated context: cosmic chronometers show a preference for the $H_0$ value derived from the CMB~\cite{2018JCAP...04..051G}, time delay and strong lensing from Quasars (QSOs) favour the $H_{0}$ from SNe Ia~\cite{2019ApJ...886L..23L}, while QSOs~\cite{Lenart:2022nip}, the Tip of the Red-Giant Branch (TRGBs)~\cite{2021ApJ...919...16F,Scolnic:2023mrv,Anderson:2023aga,Uddin:2023iob}, and Gamma-ray bursts (GRBs)~\cite{cardone09, cardone10, Dainotti2013a, postnikov14, 2021ApJ...912..150D, galaxies10010024, Dainotti:2022rfz, 2022Cao, 2022PASJDainotti,Cao2022MNRAS.510.2928C,Cao2022MNRAS.516.1386C,Dainotti2023alternative,Dainotti2023MNRAS.518.2201D} hint at an intermediate value of $H_0$ between the one of the CMB and the $H_0$ of the SNe Ia. 
To solve this intriguing puzzle and test whether the flat $\Lambda$CDM model still represents the most suitable description of the Universe, reliable cosmological probes are required at redshifts between z = 2.26 and z = 1100. To date, the best candidates for this purpose are GRBs and QSOs.  
 
In this framework, QSOs have recently attracted more and more interest among the cosmological community~\cite{rl19, 2021A&A...649A..65B, Lenart:2022nip,  2022MNRAS.515.1795B, Wang:2022hko, DainottiQSO, Li:2022inq, Pourojaghi:2022zrh,Zaja2023arXiv230508179Z} since they are observed up to z = 7.642~\cite{2021ApJ...907L...1W}, at redshifts much higher than the maximum redshift at which SNe Ia are observed. The most common efforts to standardise QSOs as cosmological candles are based on the Risaliti-Lusso (RL) relation between the logarithms of the Ultraviolet (UV) luminosity at 2500 \AA, $L_{UV}$, and the X-ray luminosity at two keV, $L_X$.


This empirical relation has been validated with several QSO samples~\cite{steffen06, just07, 2010A&A...512A..34L, lr16, 2021A&A...655A.109B}. It has been turned into a cosmological tool by carefully selecting the QSO sources to remove observational biases~\cite{rl19, lr16, rl15, salvestrini2019, 2020A&A...642A.150L}. This relation between UV and X-ray luminosities is also theoretically supported by the most accredited QSO model in which an accretion disk powers the central supermassive black hole, converting mass into energy (see, e.g. \cite{Srianand1998A&A...334...39S, Horowitz_1999,Netzerqsophysics,Kroupa2020MNRAS.497...37K}). In this scenario, the UV emission of the accretion disk is then reprocessed in X-rays from an external region of relativistic electrons via the inverse Compton effect (see, e.g.,\cite{2023ApJ...958..126F, 2024MNRAS.527.1674F, 2024MNRAS.527.1884F}). Nevertheless, this mechanism still requires a comprehensive understanding to account for the stability of the X-ray emission, which is unexpected, given that electrons are expected to cool down and fall on the central region. Thus, this stability requires an efficient energy transfer between the central and external areas, whose origin is yet to be unveiled ~\cite{2017A&A...602A..79L}. The parameters of this relation determine the amount of energy transfer from the X-ray to UV bands. 
Besides, ~\cite{DainottiQSO} has confirmed the reliability of the RL relation in cosmology, proving that, although it undergoes redshift evolution and selection effects, such as the Malmquist bias, the correction for the evolution confirms the correlation slope. Thus, this relation is entirely intrinsic to the QSO physics and not induced by selection effects or redshift evolution. This is a crucial turning point for using QSOs as cosmological tools.  

In recent years, several studies utilised the RL relation to identifying QSOs as potential cosmological standard candles~\cite{Lenart:2022nip, rl19, 2022MNRAS.515.1795B, Wang:2022hko, Li:2022inq, 2020MNRAS.497..263K}. 
Usually, they are used in conjunction with other cosmological probes since the intrinsic dispersion of the RL relation ($\delta_{int}$ $\sim$ 0.22) still limits their power in constraining cosmological parameters, compared to the precision of other probes, such as SNe Ia. For this reason, we here focus on the scavenger hunt of a sub-sample of QSOs, a `gold' sample, which presents the optimal compromise between reduced intrinsic dispersion and a sufficient number of sources to be used as a standalone probe and estimates the values of $\Omega_{M}$. We follow a similar approach to the one of finding a standard candle in the GRB domain~\cite{Dainotti2008, Dainotti:2010ki,dainotti11a,Dainotti2013a,dainotti15,dainotti2015b,2016ApJ...825L..20D,dainotti17a,dainotti2020a, Dainotti2020b,Dainotti:2021pqg,2022PASJDainotti, Dainotti:2010ki}, leading to the definition of the `Platinum' GRB sample, which has been used in several cosmological analyses~\cite{Dainotti:2022rfz,Dainotti:2022bzg}. Indeed, establishing the morphological qualities of the light curves and their spectral features is essential for selecting a standard candle. In the case of GRBs, the morphological feature that drives a standard candle is the plateau emission with peculiar characteristics.
Regarding QSOs, the key feature under consideration here is the obedience to a linear relation between the logarithmic fluxes ($F_{X} - F_{UV}$). This relation is the observer frame relation corresponding to the established RL luminosity ($L_{X} - L_{UV}$) relation in the rest frame. To achieve this standard set, starting from the initial QSO sample, several QSOs have been discarded by investigating different features. This procedure allows the sample to present well-defined properties, not to be hampered by a low signal-to-noise ratio, and not to be severely affected by extinction, UV reddening, and contamination of the host galaxy. Creating a reliable standard candle requires a deep understanding of the selection biases and redshift evolution. Moreover, the correlation and its coefficient used to calculate the distance has to be proven to be distance-independent. The most widely used method is the so-called binning procedure, where one divides the sample according to the distance of the sources and fits the correlation independently in each bin. A recent example which studies the RL correlation is~\cite{Khadka2023}. However, this procedure suffers from the fact that the borders on the bins are chosen arbitrarily. Here, we carefully study whether binning results are representative of the whole sample and have proposed a few criteria that allow us to divide the sample in a less arbitrary way.
The presented analysis aims to show a reliable methodology that determines a `gold' sample of the QSO for cosmological studies as its last goal. This work is an improvement of the work in~\cite{Dainotti2023ApJ...950...45D} because here we have performed a binned a $\sigma$ clipping procedure mutated by \citep{lusso2019}, but with a refined criterion for the analysis using robust regression methods for the outlier removal, such as the Theil-Sen regressor. This methodological approach results in a probability distribution for the values of $\Omega_M$, with the mean values centered around $\Omega_M=0.240 \pm 0.064$.
The aim of this paper can be summarised in three points:
\begin{enumerate}
    \item Create the least arbitrary binning division possible.
    \item Utilise the binning and $\sigma$ clipping procedures to pinpoint the `gold' sample, composed solely of sources that better fit the RL relation.
    \item Perform a cosmological computation using the obtained `gold' sample and check to the extent the uncertainty on $\Omega_M$ can be minimised.
\end{enumerate}

The paper is structured as follows: in Sec. \ref{sample}, we detail the data sample and the selection criteria. In Sec. \ref{methods}, we describe the methodology to identify the golden sample. We present the `gold' sample and its properties in Sec. \ref{golden sample}. In Sec. \ref{EPcorrection}, we discuss the correction for evolution; in Sec. \ref{cosmology}, we describe the cosmological results obtained with the `gold' sample and the independence of our results from the initial values of $H_0$. In Sec. \ref{MG}, we provide a physical framework to interpret the $\Omega_M$ behaviour with a metric as a function of the scalar field $f(R)$. In sec. \ref{cosmologyf(R)}, we have applied our `gold' sample with the new definition of the distance luminosity modified by the $f(R)$ theory of gravity. In Sec. \ref{conclusions}, we provide a summary and conclusions.


\section{Methodology}\label{methods}
Here, we detail the full method, starting from the description of the data sample, then the selection of the `gold' sample, and of the robust regressor method used.
\subsection{The Data Sample}\label{sample}
Our initial QSO dataset for analysis is the most recently released one, intended for cosmological applications~\cite{2020A&A...642A.150L}. It includes 2421 sources with redshifts spanning from z = 0.009 to z = 7.54~\cite{banados2018} collected from eight different catalogues in the literature~\cite{salvestrini2019, 2019A&A...632A.109N, 2019A&A...630A.118V} and archives~\cite{2016yCat..74570110M, 2018A&A...613A..51P, 2020A&A...641A.136W, 2010ApJS..189...37E}. Additionally, a sub-sample of low-redshift QSOs is included, featuring UV observations from the International Ultraviolet Explorer and X-ray data from archives. To obtain this QSO sample suitable for cosmological analyses, possible observational biases have been thoroughly inspected and removed~\cite{rl19, lr16, rl15, salvestrini2019, 2020A&A...642A.150L}. Here, we briefly describe the steps of this selection. First, only measurements with a sufficient signal-to-noise ratio S/N$\geq 1$ are retained. Then, QSOs that manifest the presence of extinction (i.e., E(B-V)$>$ 0.1) are removed to account for UV reddening and contamination of the host galaxy. For details about these cuts in the sample, see~\cite{2006AJ....131.2766R}. The contribution of absorption in X-ray has also been removed by imposing $\Gamma_X + \Delta\Gamma_X \geq 1.7$ and $\Gamma_X\leq 2.8$ if $z< 4$ and $\Gamma_X\geq 1.7$ if $z\geq 4$, with $\Gamma_X$ and $\Delta\Gamma_X$, being the photon index and its uncertainty, respectively. Ultimately, the final sample is corrected for the Malmquist bias effect requiring $logF_{\rm X,exp}-logF _{\rm min} \geq \quad \mathcal{F}$, where $F_{X, exp}$ is the X-ray flux computed from the flux in UV by imposing the RL relation and assuming $\Omega_{M} = 0.3$ and $H_0 = 70 \rm km\,s^{-1}\, Mpc^{-1}$ in a flat $\Lambda CDM$ model. $F_{min}$ is the minimum observable flux computed for each source from the time of observation \cite{lr16, 2001A&A...365L..51W}. $\quad\mathcal{F}$ is the threshold value which is fixed to $\quad\mathcal{F}= 0.9$ for QSOs coming from the cross-match of the Sloan Digital Sky Survey Data Release 14 (SDSS DR14)~\cite{SDSS:2017oqu} with 4XMM Newton~\cite{Webb:2020rgy} or with XXL~\cite{Menzel:2015sgw}, and to $\quad\mathcal{F} = 0.5$ for the ones with measurements from the SDSS DR14 and Chandra~\cite{Evans:2010ye}. To reduce the effects of the X-ray variability, if a source has more than one X-ray observation after this selection, these observations are averaged. In our study, we analyze the complete sample of 2421 sources without any additional selection, such as the redshift cut at z = 0.7 previously used in some works~\cite{2022MNRAS.515.1795B, 2020A&A...642A.150L}. This procedure is done to prevent any possible induced bias that might arise from reducing the redshift range of the sample ~\cite{Lenart:2022nip, DainottiQSO}.

\subsection{Method for the selection of the QSO `gold' sample}
\begin{figure}
\centering
\begin{tikzpicture}[node distance={17mm}, thick, main/.style = {draw, rectangle, align=center, text width=4cm}] 
\node[main] (-1) [text width=4cm] {Initial total sample of 2421 sources}; 
\node[main] (0) [below of=-1, text width=4cm, node distance=2.5cm] {Number of bins is determined by maximizing the sources from the highest to the lowest redshift bin. The minimum number in the bin is min(N)=15}; 
\node[main] (-2) [right of=0, text width=6.5cm, node distance=6cm,fill=yellow!30] {{\bf Definitions}\\ \begin{itemize}
\item $\mathbf{\delta_{D_{\rm L}}=\log_{10} (D_{\rm L,\,max})- \log_{10} (D_{\rm L,\,min})}$.\\ \item$\mathbf{\delta_{int}}$ - intrinsic scatter of the trimmed bin.\\ \item {\bf AD(bin,trimmed bin)} - the Anderson-Darling test between the parent bin and the ones without outliers.\\ \item {\bf The `untouched' sample} - the sample of sources for which it was impossible to create a small bin because  $\delta_{D_{L}}>\delta_{int}$. When we arrive at the lowest z in our sample, and this issue occurs, we move a whole bin to the untouched since it is no longer possible to create a bin with N=min(N)
\end{itemize}}; 
\node[main] (1) [below of=0, text width=4cm, node distance=2.5cm] {bin = N farthest sources in sample}; 
\node[main] (2) [below of=1, text width=3cm, node distance=1.5cm] {Fit ($F_{X}-F_{UV}$)}; 
\node[main] (3) [below of=2, text width=4cm, node distance=1.3cm] {Trim the samples in the bins within $\sigma_{clipping}$}; 
\node[main] (4) [below of=3, text width=4cm, node distance=1.5cm] {Obtain $\delta_{D_{L}}$, $\delta_{int}$, AD(bin, trimmed bin)}; 
\node[main] (5) [below of=4, text width=5cm, node distance=1.5cm] {Is $\delta_{D_{L}}\leq\delta_{int}$ $\& \, AD>0.05$?}; 
\node[main] (6) [below left of=5, text width=0.75cm,fill=green] {Yes}; 
\node[main] (7) [below right of=5, text width=0.75cm,fill=red] {No}; 
\node[main] (8) [below left of=6, text width=4cm, node distance=2.3cm] {We restart the procedure to obtain a bigger bin: N=N+1}; 
\node[main] (9) [below right of=7, text width=4.8cm, node distance=2cm] {Is the condition fulfilled for the bin of size $N> min(N)$?}; 
\node[main] (10) [below left of=9, text width=0.75cm, node distance=2.8cm,fill=green] {Yes}; 
\node[main] (11) [below right of=9, text width=0.75cm, node distance=2cm,fill=red] {No}; 
\node[main] (12) [left of=10, text width=4cm, node distance=3cm] {Store the trimmed bin fulfilling the conditions with maximum N in the final high-quality set};
\node[main] (13) [below of=12, node distance=2.2cm, text width=4.3cm] {We move to analyze the next redshift bin. The minimum z of the current bin is the upper z-limit of the next bin.}; 
\node[main] (125) [below left of=13, text width=3cm, node distance=2.7cm] {Our high-quality set will be used for cosmology};
\node[main] (14) [below of=11, text width=4cm, node distance=1.5cm] {Is the current bin at the lowest redshift possible?}; 
\node[main] (15) [below left of=14, text width=0.75cm, node distance=2cm,fill=green] {Yes}; 
\node[main] (16) [below right of=14, text width=0.75cm, node distance=2cm,fill=red] {No}; 
\node[main] (17) [below of=16, text width=4cm, node distance=2cm] {Move one source with the highest z from a whole sample to the 'untouched' sample}; 
\node[main] (18) [below left of=15, text width=3cm, node distance=2.5cm] {Move whole bin to the 'untouched' sample};

\draw[->] (-1) -- (0); 
\draw[->] (0) -- (1); 
\draw[->] (1) -- (2); 
\draw[->] (2) -- (3); 
\draw[->] (3) -- (4); 
\draw[->] (4) -- (5); 
\draw[->] (5) -- (6); 
\draw[->] (5) -- (7); 
\draw[->] (6) -- (8); 
\draw[->] (8) to [out=135,in=180,looseness=0.5] (1); 
\draw[->] (7) -- (9); 
\draw[->] (9) -- (10); 
\draw[->] (9) -- (11); 
\draw[->] (10) -- (12); 
\draw[->] (12) -- (13); 
\draw[->] (13) -- (125); 
\draw[->] (13) to [out=160,in=170,looseness=0.5] (1); 
\draw[->] (11) -- (14);
\draw[->] (14) -- (15); 
\draw[->] (14) -- (16); 
\draw[->] (15) -- (18); 
\draw[->] (16) -- (17); 
\draw[->] (17) to [out=50,in=335,looseness=0.6] (1); 

\end{tikzpicture} 
\caption{The flowchart of our algorithm.}
\label{fig:algorithm}
\end{figure}
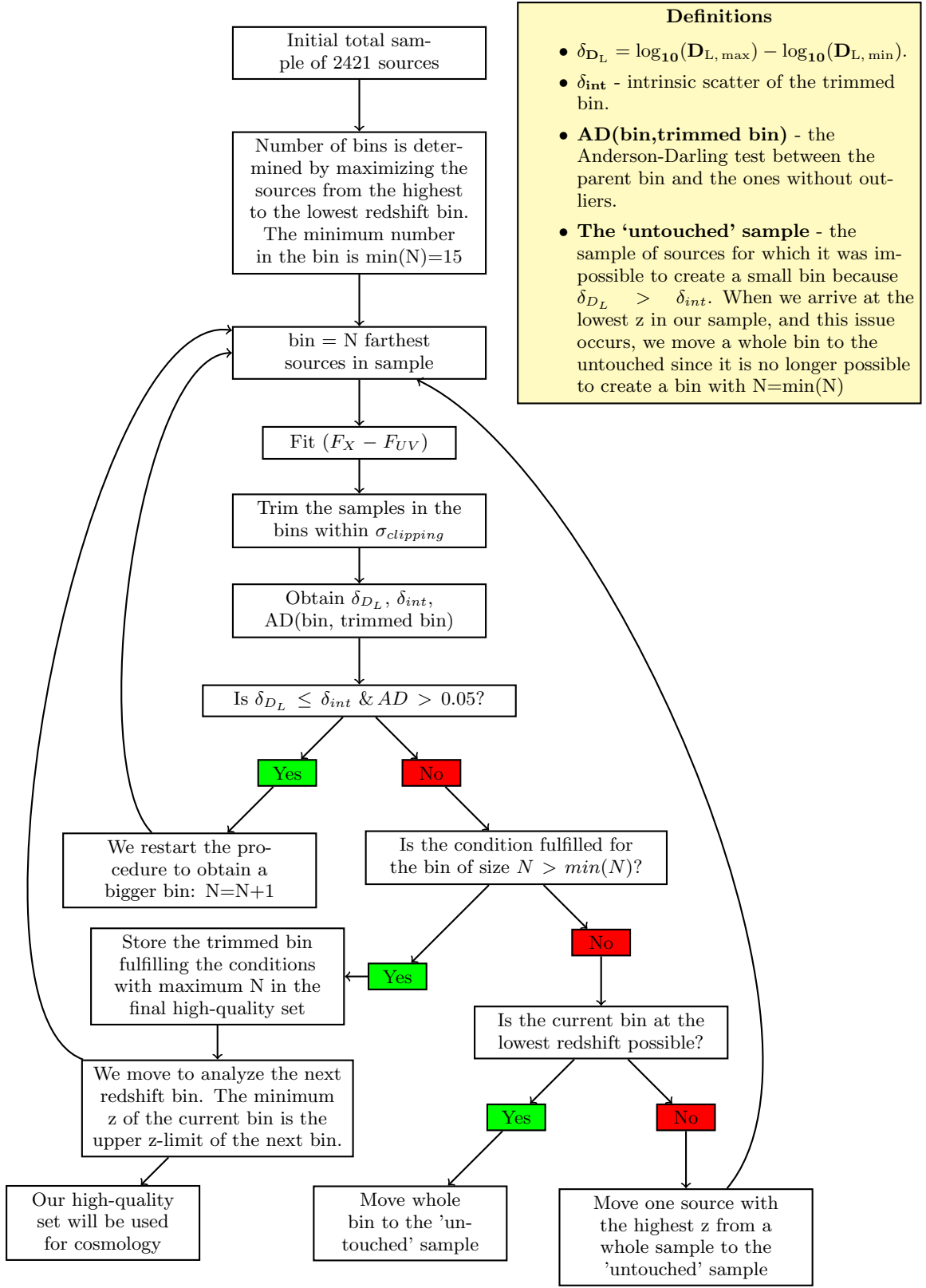

For clarity of the reader, we have added a flowchart to detail the steps of the procedure; see Fig \ref{fig:algorithm}.

Here, we detail the procedure applied to select the `gold' QSO sub-sample. We first need to start from the $L_X-L_{UV}$ relation. The functional form of this relation is as follows:

\begin{equation}
  \log_{10}L_X = \gamma \times \log_{10}L_{UV}+{\beta}.
  \label{lumcor}
\end{equation}
where $\gamma$ is the normalization and $\beta$ is the coefficient that determines the amount of energy transfer from the X-ray to UV bands.
More specifically, $L_X$ and $L_{UV}$ correspond to the X-ray and UV luminosities, respectively, which are derived from measured energy fluxes in given bands and are not corrected for selection biases.

\begin{equation}
  \log_{10}(F_X\, 4 \pi D_L^{2}) = \gamma \times \log_{10}(F_{UV}\, 4 \pi D_L^{2}) +{\beta},
  \label{eq:fluxes}
\end{equation}
where $\log_{10} (F_{X})$, and $\log_{10} (F_{UV})$ are the logarithms of the measured fluxes of the original QSO sample composed of 2421 sources, and $D_L$ is the luminosity distance, in our work computed for flat $\Lambda CDM$ model \citep{2016A&A...594A..13P}, which is estimated by

\begin{equation}
D_{L} = (1+z) \frac{c}{H_{0}} \int_{0}^{z} \frac{dz'}{\sqrt{\Omega_{M} (1+z')^3 + (1-\Omega_{M})}},
\label{distance_Lum}
\end{equation}
with the Hubble constant $H_0=70\,{\rm km\,s^{-1}\, Mpc^{-1}}$ as a reference value for the standard sample of QSOs, 
with $\Omega_{M}$ being the matter density parameter and left free to vary, and $c$ is the speed of light. After performing some simple algebra on the Eq. \ref{eq:fluxes}, one obtains:

\begin{equation}
    log_{10}(F_{X})=\gamma \times log_{10}(F_{UV}) + (\gamma -1) \times log_{10}(4\pi D_L^2) + \beta.
\end{equation}

When we examine only QSOs that are closely spaced in redshift bins, the expression $(\gamma -1)\times log_{10}(4\pi D_L^2) + \beta$ can be considered to be constant. Therefore, in the small bin, the linear correlation between fluxes, expressed as $log_{10}(F_{X})=\gamma \times log_{10}(F_{UV}) + \beta_{flux}$, continues to be a valid estimate for the linear correlation in luminosities with different normalizations.


We define a bin as small when the intrinsic scatter, denoted as $\delta_{int}$, dominates over the dispersion caused by the difference in the distances in each bin. The extent to which the correlation is influenced by the difference in distances is represented by the formula $|\gamma - 1| (log_{10}(D_{L,\, max}) - log_{10}(D_{L,\, min}))$. In this context, the factor $|\gamma - 1|$ is always smaller than 1. Thus, the condition $\delta_{int} > (log_{10}(D_{L,\, max}) - log_{10}(D_{L,\, min}))$ weighs more (resulting in tighter bins in redshift) than the same condition modified by the $|\gamma - 1|$ factor. Ultimately, the size of our bin is determined by the threshold we set for the 4 $\sigma$-clipping, which in turn dictates the number of outliers removed. The greater the number of outliers removed, the smaller the intrinsic scatter and the smaller the allowed difference in distance. Therefore, in practice, by establishing a range of thresholds for $\sigma_{clipping}$, we are able to investigate a large number of bin sizes.
We further define a small bin with the following condition:

\begin{equation}
    \delta_{D_{\rm L}}=\log_{10} (D_{\rm L,\,max})- \log_{10} (D_{\rm L,\,min}) \leq \delta_{\rm int}.
    \label{condition}
\end{equation}
Consequently, the influence of $\delta_{D_{L}}$ is relatively minor, and its contribution is less significant than the dispersion of the relation itself. Therefore, the differences in the luminosity distances, and by extension, the associated cosmological model, can be considered negligible in comparison to the scatter of the relation.



Furthermore, in our analysis, in parallel, we divide the sample into bins and apply the $\sigma$ clipping procedure. To describe the $\sigma$ clipping method, we define the distance of a point from the best-fit line as:

\begin{equation}
    y=\sigma-clipping=\left|\frac{log_{10}(F_X)-\gamma\times log_{10}(F_{\rm UV})-\beta}{\sqrt{\sigma^{2}_{log_{10}(F_X)}+\gamma^2\times \sigma^{2}_{log_{10}(F_{UV})}+\delta^2_{int}}}\right|,
\end{equation}
where $\sigma_{log_{10}(F_X)}$ and $\sigma_{log_{10}(F_{UV})}$ are the uncertainties on the logarithmic fluxes in X-ray and UV bands, respectively. Thus, the $\sigma$-clipping is the error-normalised difference between the observed and predicted X-ray flux $F_X$ in the RL relation. It is computed independently for each point.
Then, we decide a given distance $\sigma_{clipping}$), and all sources with $y<\sigma_{clipping}$ are considered `inliers'. The smaller the value chosen for the $\sigma$ clipping, the closer the data points are to the fitting line, thereby reducing dispersion around the studied relation.
Thus, the dispersion of the relation is always relatively small, which makes the condition \ref{condition} mentioned above even more constraining in terms of the difference in distance.

We defined the condition that the bin has to be considered `small', but we also aim to maximize the number of sources in a given bin. Furthermore, our scope is to use QSOs at very high redshifts to minimise the impact of having a few QSOs at high-z. Indeed, the high redshift sources are the ones more interesting for us for building the standard set of QSOs. Thus, we initiate the binning division by creating an initial sample at the highest redshift.
This initial sample is composed of the largest number of data points possible while still adhering to the criteria in Eq. \ref{condition}.
After we have established the highest redshift bin division, we continue with creating smaller redshift bins until we cover the entire redshift range. Before we detail the procedure, it is essential to clarify the terminology used. The term `untouched' refers to sources that do not meet the criteria set out in Eq. \ref{condition} as a result of the $\sigma$ clipping process. Consequently, these sources should remain untouched; they are not suitable for use in the $\sigma$ clipping procedure. `untrimmed' sources, on the other hand, are those for which the $\sigma$ clipping process has not yet been applied, and they remain part of the original bin sample. Finally, the `trimmed' sources are those that have undergone the $\sigma$ clipping process, with outliers having been removed from the initial sample.
A flowchart in Fig. 1 graphically represents the basic steps of our algorithm.
We provide a detailed description of the procedure below.

\begin{enumerate}
    \item Initially, we selected a minimum number of 15 sources with the highest redshifts in our dataset. The number of 15 sources is the smallest possible to obtain a sample of at least 3 sources after trimming in the cases where we discard the highest number of outliers. The number of sources in a given bin before the $\sigma$ clipping varies in our results from 15 to 487. Indeed, we here stress that this procedure is independent of the initial set we choose for the number of sources in a bin because the conditions detailed below will set the final number.
    \item We perform the Theil-Sen linear 2D fitting of the $log_{10}(F_{X})-log_{10}(F_{UV})$ correlation for each bin. This is a robust and distance-independent method described in \ref{theilsen}.
    \item We utilise the results of this fitting to apply the $\sigma$ clipping procedure to the dataset within each given bin.
    \item To guarantee that the original sample before the $\sigma$ clipping is drawn from the same parent population of the initial sample, we applied the two-sample Anderson-Darling (AD) test between these sets. 
    \item We ensure the condition \ref{condition} for each bin.
    \begin{enumerate}           
        \item As previously stated, we apply a $\sigma$ clipping procedure that requires a minimum sample size of 15 objects. When a sample of 15 does not satisfy the condition \ref{condition}, we remove one source from the 15 at the highest redshift and place it in a sample that is called the `untouched' set.
        Subsequently, we add the one source we had previously not included at low redshift and proceed to perform the procedure outlined in step 2.
        At these lower redshifts, we have a greater number of data points, resulting in smaller distances between sources. 
        \item In all other cases, we continue with the steps below.
        \end{enumerate}
        \item We gradually add sources at smaller redshifts to enlarge our sample. Thus, if the set of 15 sources at the highest redshift, after the trimming, fulfills the condition \ref{condition}, we test a sample of 16 sources at the highest redshift. We continue with 17, 18, etc. until the condition \ref{condition} is still fulfilled for that particular $\sigma$ clipping.
    This way, we probe the behaviour of all the possible sizes of a given bin around the point $\delta_{D_{L}}=\delta_{int}$, see the left panel of Fig. 2.  
    
    \item We choose the Theil-Sen method over RANSAC because the residuals from the best fit are not always Gaussian, whereas RANSAC has a Gaussianity requirement.
       
    \item Through this process, we obtain several outputs: the untrimmed set of sources, the trimmed sources, their scatter, and the $\delta_{D_{L}}$ for this set, as well as the p-value from the AD test between the trimmed and non-trimmed sets.
    We choose the most numerous set of sources that satisfies all the specified conditions: $\rm p-value>0.05$, $\delta_{D_{L}}<\delta_{int}$ and the number of QSOs in the trimmed set $\geq 3$. 
    We store all the sources that are considered inliers for the Theil-Sen method.
    Subsequently, we exclude the obtained untrimmed set from the entire sample and initiate a new bin analysis, starting from step 1 to derive another sample.
\end{enumerate}

\begin{figure}
    \centering
    \includegraphics[width=0.49\textwidth]{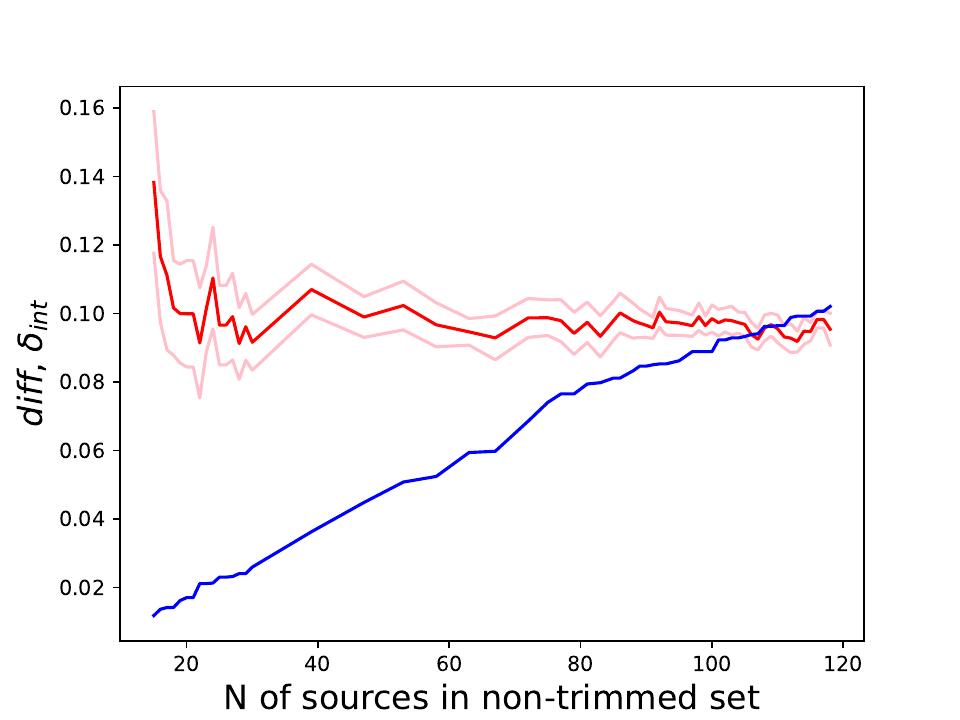}
    \includegraphics[width=0.49\textwidth]{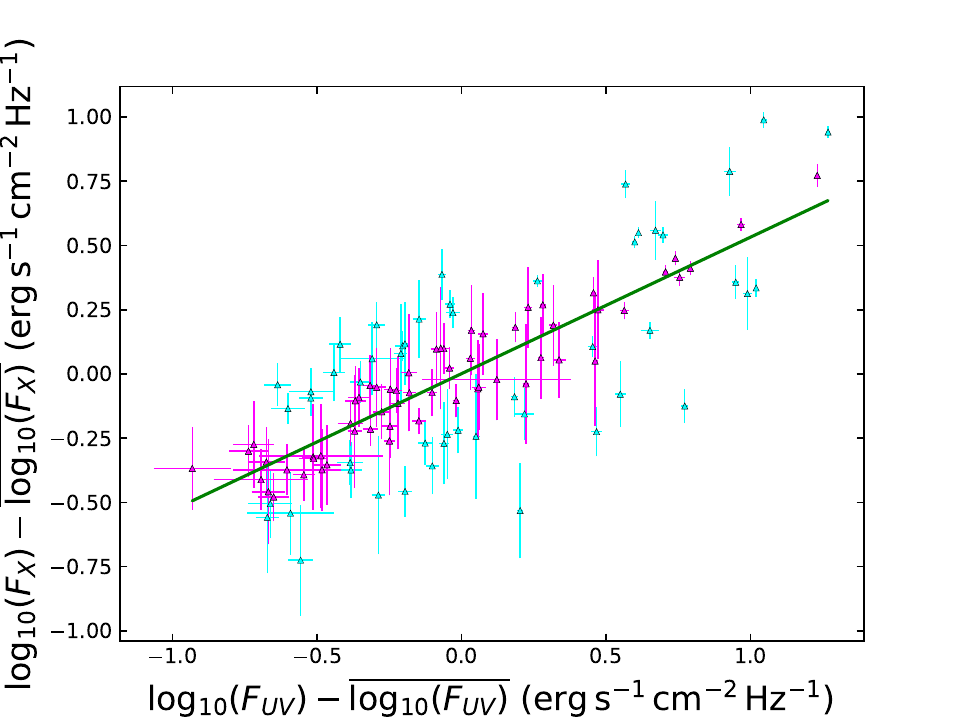}
    \caption{This figure presents the results of the applied binning algorithm. On the left panel, we illustrate how the gradual addition of sources to the set changes the parameter $\delta_{D_{L}}=log_{10}(D_{L,\, max})-log_{10}(D_{L,\, min})$ of a trimmed set (blue line) and the scatter of the correlation in the trimmed set $\delta_{int}$ (red line). The pink lines represent the 1 $\sigma$ error bars for the scatter. On the right panel, we present the flux-flux distribution within the obtained bin. Sources that are in agreement with the best-fit line (shown in green) within a specific range are highlighted in purple. Cyan represents sources within the bin that are considered outliers. The above plots were generated for the fourth, the most distant redshift bin obtained for the $\sigma-clipping=0.6$.}
    \label{diff_sv}
\end{figure}

We stress that it is a rare occurrence to have only 3 sources in a bin after we remove the outliers with the $\sigma$ clipping method. This only happens for one bin when we use $\sigma$ clipping interval of $0.2$. For our best sample, which uses a $\sigma$ clipping interval of $0.6$, the smallest number of sources in a bin is 7.
For each bin, after we obtain the trimmed set of sources, we combine them with the set of sources that have not been used for the $\sigma$ clipping method (the untouched set). This gives us our final sample for each $\sigma$ clipping interval.
We show an example of how the parameters $\delta_{int}$ and $\delta_{D_{L}}$ change with the number of sources using the interval 0.6 in the left panel of Fig. \ref{diff_sv}.
In the right panel, we present the normalised $log_{10}(F_{X})-log_{10}(F_{UV})$ relation for the sources in this bin. In the figure, sources are colour-coded according to their proximity to the best-fit line (shown in green). Those closely aligned with the best-fit line within a particular $\sigma$ clipping threshold are highlighted in purple. On the other hand, the cyan indicates sources within the bin that are outliers according to the Theil-Sen method. This presentation provides a thorough examination of the relationship between sources inside the bin and the established best-fit line. 
We show how our bins are populated before and after discarding outliers at the left and right panels of Fig. \ref{bins_size}, respectively. The different colors correspond to different $\sigma_{clipping}$ values used in our algorithm.
It is clear from Fig.~\ref{bins_size} that the maximum number of sources is obtained in the redshift range between 1 and 2, as we have a large population of sources and this number decreases as the number of the $\sigma_{clipping}$ decreases. Indeed, the higher the $\sigma_{clipping}$ value, the fewer sources are considered outliers in the procedure. The situation is similar also after we discard the outliers, as we can see from the right panel of Fig. \ref{bins_size}.
Fig. \ref{bins_numb} shows the number of bins as a function of $\sigma_{clipping}$ values color-coded on the right side according to the number of QSO sources present in each bin for a given $\sigma_{clipping}$ value. From this figure, it is evident that the smaller the value of $\sigma_{clipping}$, the larger the number of bins, as there will be fewer sources in each bin. This occurs because the data must be closer to the fitting line while simultaneously fulfilling the condition of $\delta_{DL} < \delta_{int}$, achievable only with bins restricted to smaller redshift intervals. Conversely, the larger the $\sigma_{clipping}$ value, the more sparsely the sources are distributed around the best-fit line. Thus, we can afford a smaller number of bins since the $\delta_{DL} < \delta_{int}$ is reached with larger bins in redshifts.

\begin{figure}[ht!]
    \centering
    \includegraphics[width=0.49\textwidth]{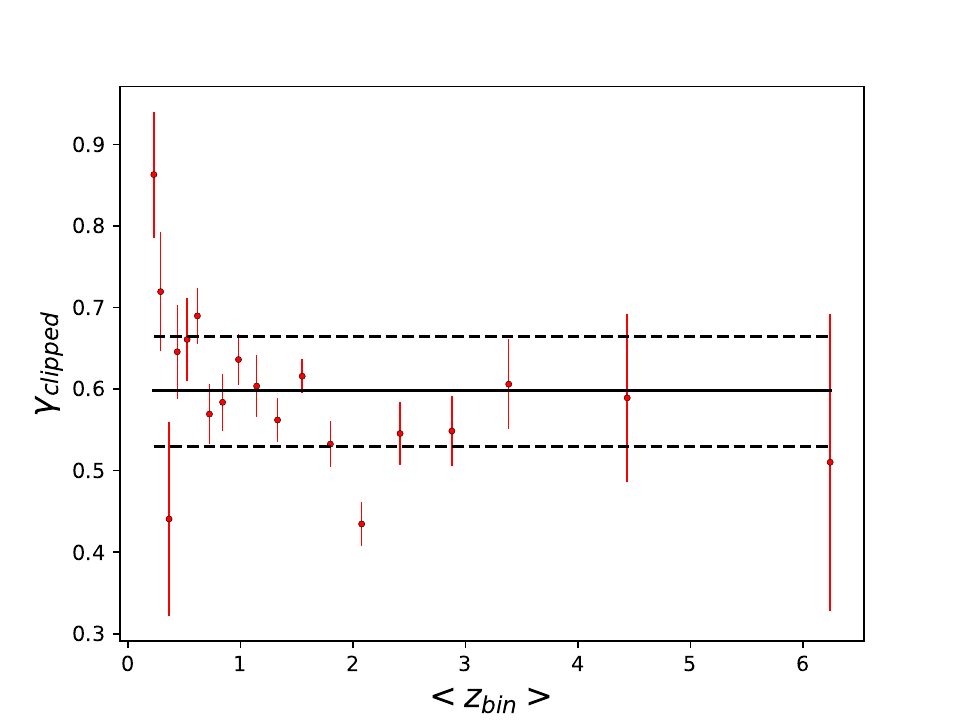}
    \includegraphics[width=0.49\textwidth]{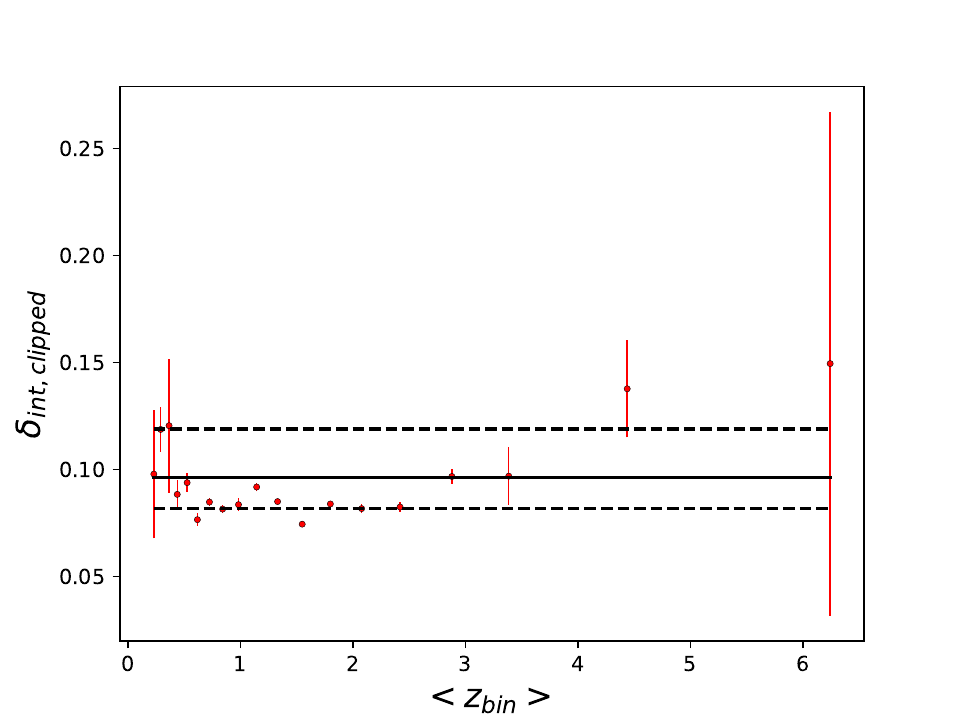}
    \caption{The left panel shows the relationship between the slope, denoted as $\gamma$, and the mean redshift $<z_{bin}>$ for each bin within the sample. In the right panel, we show the intrinsic dispersion $\delta_{int}$ within each bin versus the mean redshift. The error bars in both panels correspond to the 68\% confidence level intervals (CLI), which help to assess the reliability of our fitting. These panels are specific to the interval 0.6 in the $\sigma$ clipping. The solid bold line corresponds to the mean value of slope (left) and dispersion (right) from all bins. The dashed lines correspond to the 68\% CLI of the obtained parameter distributions.}
    \label{bins_slope}
\end{figure}

\begin{figure}[ht!]
    \centering
    \includegraphics[width=0.49\textwidth]{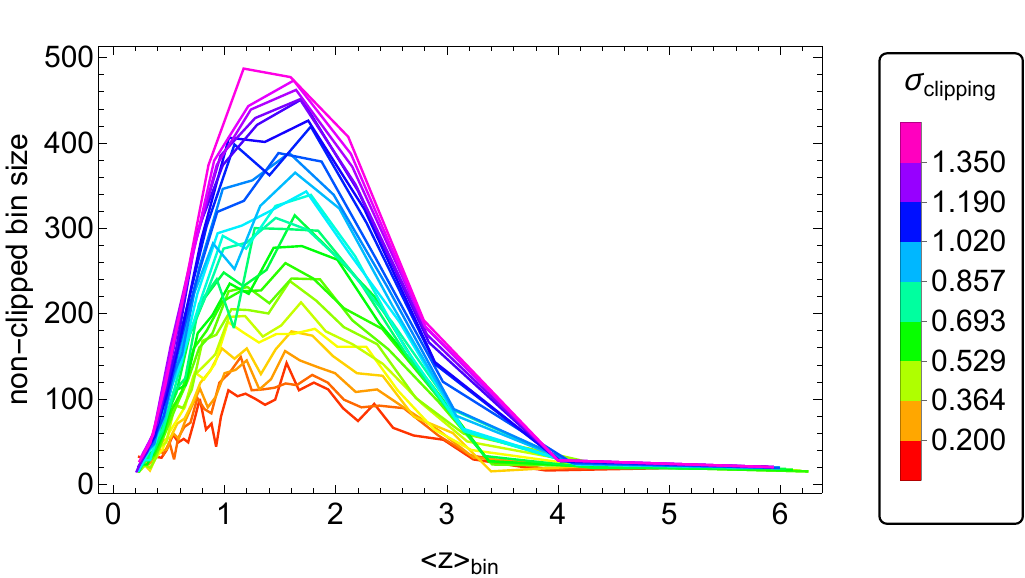}
    \includegraphics[width=0.49\textwidth]{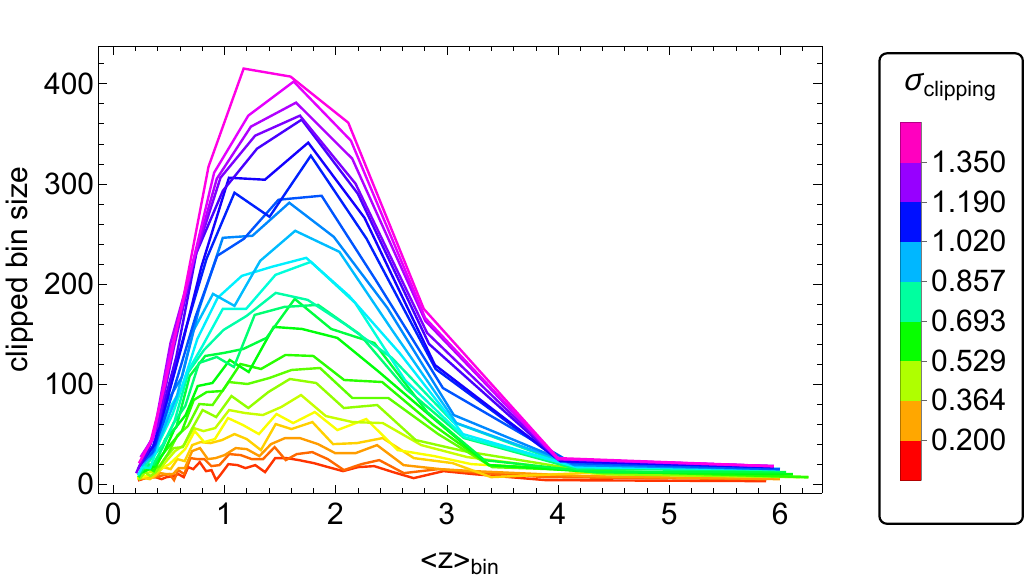}
    \caption{The curves show the number of data points in each bin vs the mean redshift of a given bin. Different colors of the curves correspond to different $\sigma_{clipping}$ used in the binning procedure. On the left panel, the sizes of the bins are shown before the $\sigma$ clipping, while on the right, they are shown after outliers have been removed.}
    \label{bins_size}
\end{figure}

\begin{figure}[ht!]
    \centering
    \includegraphics[width=0.95\textwidth]{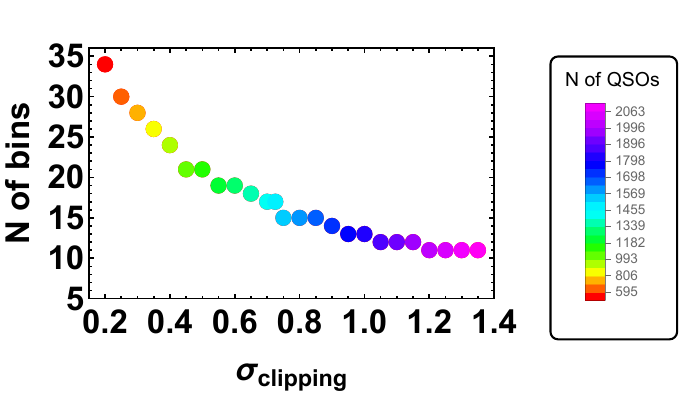}
    \caption{Number of bins obtained in our procedure as a function of the $\sigma$-clipping color-coded according to the number of sources contained in each bin for a given $\sigma$- clipping.}
    \label{bins_numb}
\end{figure}

As previously discussed, we approximate the RL relation in each bin with the $\log_{10}(F_X)-\log_{10}(F_{UV})$ relation. 
The slopes $\gamma$ of all bins and their dispersion $\delta_{int}$ are compared with each other on the left and right panels of Fig. \ref{bins_slope}, respectively.
As evident from Fig. \ref{bins_slope}, all slopes in separate bins align consistently with the mean value. All values are compatible with the mean within $<2.2\sigma$. Furthermore, we find no significant correlation between the value of slope $\gamma$ and the mean redshift of the bin $<z_{bin}>$. Although the values at the smallest and highest redshift are characterised by high uncertainty, it is worth noting that the scatter $\delta_{int}$ across all the investigated bins shows very small variation in most of the studied cases, not only for the `gold' sample.

\subsection{The Thielsen method: a robust regression}
\label{theilsen}

Given that we fit the linear relation in separate bins, it's crucial to perform this fitting at the `centre' of the flux-flux space for each bin. Accordingly, we consistently fit the relation:

\begin{equation}
    \log_{10}(F_{X,\, bin})-\overline{\log_{10}(F_{X,\, bin})}=\gamma \times (\log_{10}(F_{UV,\, bin})-\overline{\log_{10}(F_{UV,\, bin})}) + \beta,
\end{equation}
where the horizontal line indicates the variable's mean value. When the simple linear fitting is performed, the slope value is usually correlated with the normalisation value. Fitting at the center of coordinates helps to eliminate this correlation. This is crucial in the binned analysis because here, $\beta$ is highly dependent on the distance. Thus, a not normalised fitting would produce an artificial correlation between the slope and distance. 
Furthermore, we employ linear regression to identify a highly accurate sample. Therefore, it is essential to use a robust fitting technique. One such method is a Theil-Sen estimator, which is insensitive to outliers and can be applied to data whose dispersion does not follow a normal distribution.

The Theil-Sen method, named after the Dutch statisticians Henri Theil and B. Gustavsen and American economist Edwin Tedeschi, differs from traditional methods by seeking the median slope from all data point pairs. It can be significantly more accurate than the least squares method when dealing with outliers and unusual data points. This approach excels in handling outliers by determining the median slope to describe the relationship between variables. This estimator relies on the chosen seed for a random number generator. Thus, we repeat this fitting 100 times and take the mean and standard deviation of the obtained results as a final value of a given parameter and its respective uncertainty.
Regarding point 7 in the previous section, it's important to note that these robust regression methods are advantageous due to their varied applicability conditions. Indeed, in principle, we could have also applied the RANSAC method.
The Random Sample Consensus (RANSAC) algorithm is a versatile approach for parameter estimation, particularly designed to handle a substantial proportion of outliers within the input data by iteratively identifying and weighing these outliers less, allowing the algorithm to focus on the more reliable data for the model estimation. RANSAC operates as a non-deterministic algorithm, providing reasonable results with a certain probability that increases with more iterations. Essentially, it can be viewed not only as a parameter estimation method but also as an outlier detection technique, emphasizing its robustness in the presence of noisy data.
However, the RANSAC method requires the Gaussianity assumption for the residuals, which is not the case in our sample. One could wonder if the trimming itself would cut the sample in a way to trim the tails of the Gaussian distribution so that it would no longer be Gaussian. At the end of our procedure, we check if the entire sample fulfills the Gaussianity condition on the residuals. Then, we need to apply a robust regression method, which, by definition, does not cut the tails of the distribution, in contrast to the Ransac method. Otherwise, we would never be sure about the statistical nature of the residuals, namely if they are Gaussian or not. 
In addition, the Theil-Sen method effectively reduces the weight of outliers, enabling a more precise fit with smaller scatter, even using the same $\sigma$-clipping parameter, compared to traditional linear regression techniques. This results in more reliable and accurate parameter estimations, particularly in datasets where outliers might otherwise skew the results \citep{Dainotti:2023cpn}.
While Theil-Sen and Ransac are robust regressions that weigh the outliers differently, they do not discard the sources, and our aim is to pinpoint a sample of higher-quality sources for cosmology. This is the reason why we combine both the regression technique and the $\sigma$-clipping approach. In addition, there are limitations to using and implementing these robust regression methods within cosmological procedures. 
The Theil-Sen estimator cannot be straightforwardly used in cosmological computations because this method works well only for a linear, two-dimensional fitting. In the cosmological computation involving the RL relation, we effectively fit a LCDM model together with 3 observables ($F_X$, $F_{UV}$, z) with a non-linear relation. To perform an outlier-independent cosmological fitting, it is necessary to first identify and remove outliers from the entire sample and then perform fitting on the cleaned data set. This necessity underpins our analysis. Our goal is to achieve model-independent data trimming; hence, we conduct it on the fluxes. Fluxes are an appropriate approximation of luminosity only in small bins, which is why we clean the data in narrow redshift ranges. To ensure that outliers do not affect the data-cleaning process, we employ a robust estimator: the Theil-Sen method. The data cleaning in our analysis is a method of discarding data by choosing a $\sigma$-clipping value higher than a given threshold. Then, we investigate a grid of thresholds to find a compromise between the scatter of the correlation and the number of points in our final sample.

\section{Results}
\subsection{The identification of the golden sample}\label{golden sample}
\begin{figure}[ht!]
  \centering  \includegraphics[width=0.95\textwidth]{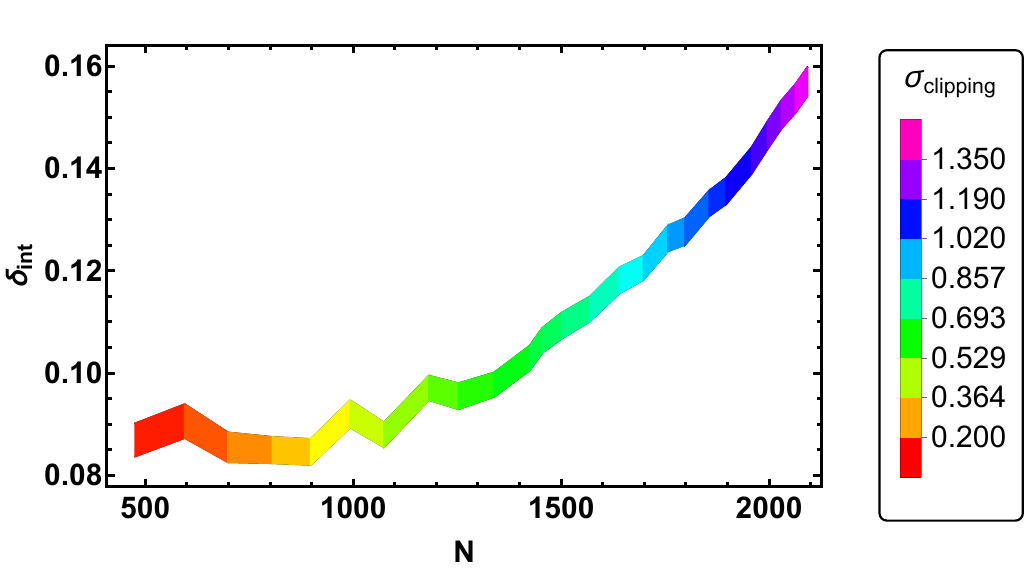}
  \caption{The plot illustrates the intrinsic dispersion, $\delta_{int}$, of the sub-samples obtained with the $\sigma$ clipping method as a function of the size of these sub-samples. The colour coding with the colour bar indicates the interval chosen in the $\sigma$ clipping algorithm. The thickness of the rainbow line indicates the range of 1 $\sigma$ error bars.}
  \label{fig:scatter}
\end{figure}

\begin{figure}[ht!]
  \centering  \includegraphics[width=0.95\textwidth]{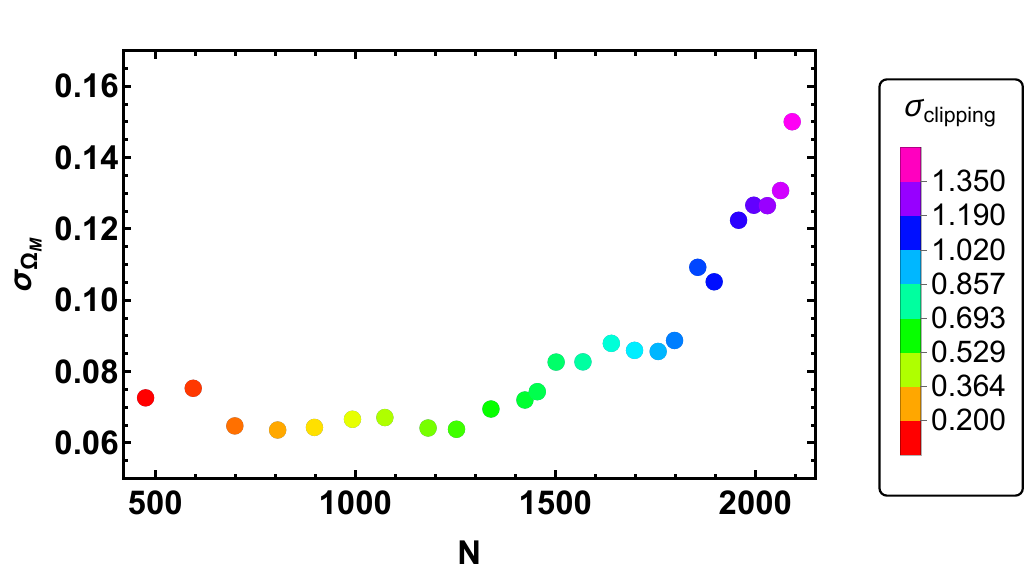}
  \caption{The matter density $\Omega_{M}$ uncertainty parameter as a function of the number of sources obtained in the $\sigma$ clipping procedure with values colour-coded in the colour bar located on the right side of the figure.}
  \label{fig:matter}
\end{figure}

\begin{figure}[ht!]
  \centering  \includegraphics[width=0.9\textwidth]{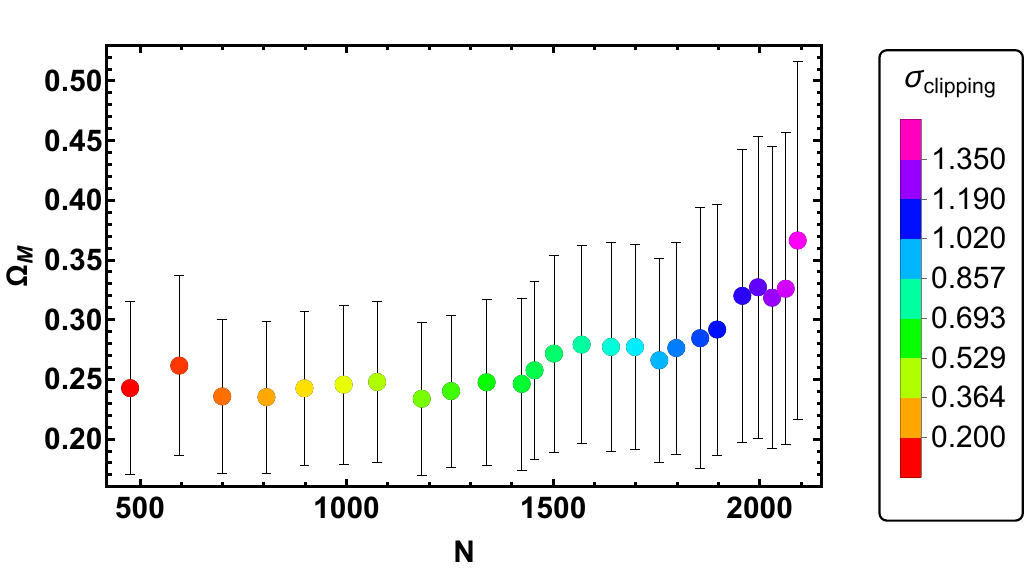}
  \caption{The matter density parameter, $\Omega_{M}$, computed for each sub-sample obtained in the $\sigma$ clipping process, is shown as a function of the size of these subsets. The colour scheme indicates the interval chosen in the $\sigma$-clipping algorithm. The error bars represent 1 $\sigma$ uncertainties.}
  \label{fig:matter3}
\end{figure}


\begin{figure}[ht!]
  \centering  \includegraphics[width=0.9\textwidth]{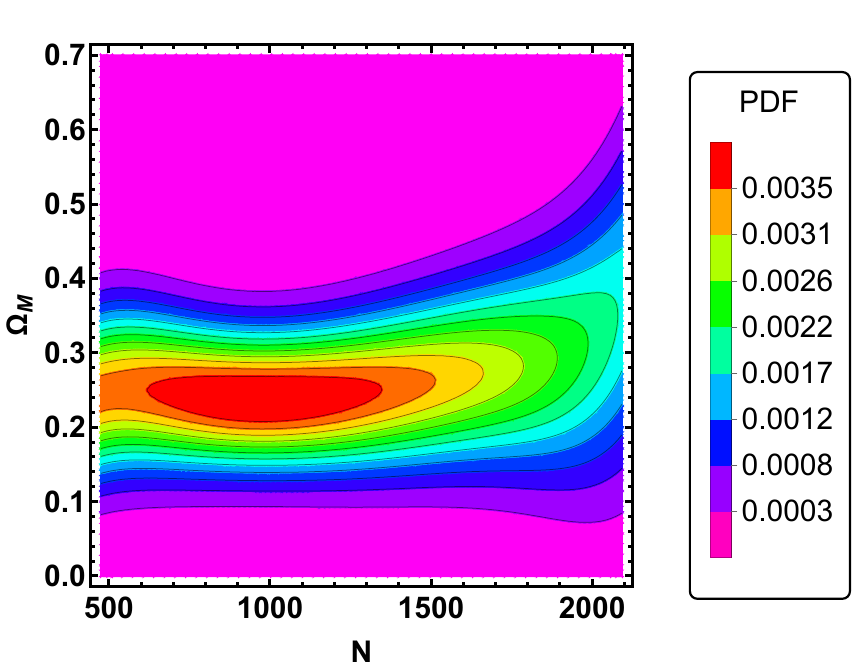}
  \caption{The matter density parameter, $\Omega_{M}$, computed for each sub-sample obtained through the $\sigma$ clipping procedure as a function of the sub-sample size. The color scale signifies the probability distribution associated with achieving the smallest scatter in $\Omega_M$.}
  \label{fig:matter2}
\end{figure}

\begin{table}
	\begin{center}
		\caption{The table provides a breakdown of columns, starting from the first, which is the number of bins, and the second is the $\sigma$-clipping values. Subsequent columns include the number of sources that survived after the $\sigma$ clipping procedure, the dispersion of the $log_{10}(L'_{X})$-$log_{10}(L'_{UV})$ correlation for this sample, as well as the values of $\Omega_M$ and their respective uncertainties fixing the values of $H_0$ at $70$, $73.5$ and $67.5$ km $s^{-1} Mpc^{-1}$. The subsequent two columns correspond to the z-score test of the value obtained in a particular row with value for the `gold' sample ($\Omega_{M}=0.240\pm 0.064$) and the SNe Ia ($\Omega_{M}=0.334\pm 0.018$), respectively. The z-score is computed as $\frac{\Omega_{M, i}-\Omega_{M, comparison}}{\sqrt{\sigma_{\Omega_{M, i}}^2+\sigma_{\Omega_{M, comparison}}^2}}$. The details are discussed in the main text.}
  \label{tab:results}
  \begin{tabular}{ |c|c||c|c|c|c|c|c|c| } 
 \hline
 \makecell{{\bf number of} \\bins} & $\sigma$ clipping & \makecell{N of survived and \\ untouched sources} & $\delta_{int}$ & $\Omega_M(H_0=70)$ & $\mathbf{\Omega_M(H_0=73.5)}$ & $\mathbf{\Omega_M(H_0=67.5)}$ & $\rm z_{Gold}$ & $\rm z_{SNe}$\\
 \hline\hline
$34$ & $0.20$ & $476$ & $0.087\pm 0.003$ & $0.243\pm 0.073$ & $0.249\pm 0.071$ & $0.248\pm 0.072$ & $0.01$ & $-0.38$  \\\hline
$30$ & $0.25$ & $595$ & $0.091\pm 0.003$ & $0.262\pm 0.075$ & $0.262\pm 0.078$ & $0.255\pm 0.078$ & $0.06$ & $-0.28$  \\\hline
$28$ & $0.30$ & $699$ & $0.086\pm 0.003$ & $0.236\pm 0.065$ & $0.230\pm 0.059$ & $0.234\pm 0.063$ & $-0.01$ & $-0.42$  \\\hline
$26$ & $0.35$ & $806$ & $0.085\pm 0.003$ & $0.235\pm 0.064$ & $0.239\pm 0.061$ & $0.239\pm 0.063$ & $-0.02$ & $-0.42$  \\\hline
$24$ & $0.40$ & $898$ & $0.085\pm 0.003$ & $0.243\pm 0.064$ & $0.237\pm 0.066$ & $0.233\pm 0.057$ & $0.01$ & $-0.38$  \\\hline
$21$ & $0.45$ & $993$ & $0.092\pm 0.003$ & $0.246\pm 0.067$ & $0.243\pm 0.070$ & $0.240\pm 0.064$ & $0.02$ & $-0.36$  \\\hline
$21$ & $0.50$ & $1074$ & $0.088\pm 0.002$ & $0.248\pm 0.067$ & $0.248\pm 0.064$ & $0.250\pm 0.066$ & $0.02$ & $-0.35$  \\\hline
$19$ & $0.55$ & $1182$ & $0.097\pm 0.002$ & $0.234\pm 0.064$ & $0.239\pm 0.070$ & $0.235\pm 0.067$ & $-0.02$ & $-0.43$  \\\hline
$19$ & $0.60$ & $1253$ & $0.096\pm 0.003$ & $0.240\pm 0.064$ & $0.231\pm 0.062$ & $0.236\pm 0.063$ & $-$ & $-0.39$  \\\hline
$18$ & $0.65$ & $1339$ & $0.098\pm 0.002$ & $0.248\pm 0.070$ & $0.250\pm 0.069$ & $0.242\pm 0.066$ & $0.02$ & $-0.35$  \\\hline
$17$ & $0.70$ & $1424$ & $0.103\pm 0.002$ & $0.246\pm 0.072$ & $0.254\pm 0.078$ & $0.246\pm 0.073$ & $0.02$ & $-0.36$  \\\hline
$17$ & $0.725$ & $1455$ & $0.106\pm 0.002$ & $0.258\pm 0.074$ & $0.266\pm 0.075$ & $0.256\pm 0.068$ & $0.05$ & $-0.30$  \\\hline
$15$ & $0.75$ & $1502$ & $0.109\pm 0.003$ & $0.272\pm 0.083$ & $0.279\pm 0.087$ & $0.273\pm 0.085$ & $0.09$ & $-0.23$  \\\hline
$15$ & $0.80$ & $1569$ & $0.113\pm 0.002$ & $0.279\pm 0.083$ & $0.276\pm 0.082$ & $0.279\pm 0.089$ & $0.11$ & $-0.20$  \\\hline
$15$ & $0.85$ & $1640$ & $0.118\pm 0.003$ & $0.277\pm 0.088$ & $0.284\pm 0.092$ & $0.283\pm 0.091$ & $0.10$ & $-0.20$  \\\hline
$14$ & $0.90$ & $1698$ & $0.121\pm 0.002$ & $0.277\pm 0.086$ & $0.275\pm 0.090$ & $0.269\pm 0.083$ & $0.10$ & $-0.20$  \\\hline
$13$ & $0.95$ & $1757$ & $0.126\pm 0.002$ & $0.266\pm 0.086$ & $0.268\pm 0.087$ & $0.264\pm 0.091$ & $0.07$ & $-0.26$  \\\hline
$13$ & $1.00$ & $1798$ & $0.128\pm 0.003$ & $0.276\pm 0.089$ & $0.280\pm 0.091$ & $0.280\pm 0.095$ & $0.10$ & $-0.21$  \\\hline
$12$ & $1.05$ & $1856$ & $0.133\pm 0.002$ & $0.285\pm 0.109$ & $0.285\pm 0.094$ & $0.295\pm 0.106$ & $0.12$ & $-0.17$  \\\hline
$12$ & $1.10$ & $1897$ & $0.136\pm 0.003$ & $0.292\pm 0.105$ & $0.291\pm 0.104$ & $0.284\pm 0.094$ & $0.14$ & $-0.14$  \\\hline
$12$ & $1.15$ & $1958$ & $0.142\pm 0.003$ & $0.320\pm 0.122$ & $0.308\pm 0.107$ & $0.316\pm 0.120$ & $0.20$ & $-0.04$  \\\hline
$11$ & $1.20$ & $1996$ & $0.146\pm 0.003$ & $0.327\pm 0.127$ & $0.333\pm 0.133$ & $0.332\pm 0.135$ & $0.21$ & $-0.02$  \\\hline
$11$ & $1.25$ & $2030$ & $0.151\pm 0.003$ & $0.318\pm 0.127$ & $0.319\pm 0.133$ & $0.320\pm 0.121$ & $0.20$ & $-0.05$  \\\hline
$11$ & $1.30$ & $2063$ & $0.154\pm 0.003$ & $0.326\pm 0.131$ & $0.319\pm 0.128$ & $0.333\pm 0.135$ & $0.21$ & $-0.03$  \\\hline
$11$ & $1.35$ & $2092$ & $0.157\pm 0.003$ & $0.366\pm 0.150$ & $0.353\pm 0.143$ & $0.349\pm 0.144$ & $0.29$ & $0.09$  \\\hline
		\end{tabular}
	\end{center}
\end{table}

Remarkably, in Fig. \ref{fig:scatter}, it is evident that the values of the dispersion $\delta_{int}$ tend to approach a value of $\sim 0.09$ asymptotically as the number of sources decreases, indicating a more trimmed sample. Notably, this precise value was determined by \cite{Sacchi2022} through a thorough study of sources at $z\approx 3$. The study by \cite{Sacchi2022} applied sample cleaning based solely on data quality and argued that the value of 0.09 can be attributed exclusively to the inclination of the jets of QSOs and their variability. Indeed, the smallest $\delta_{int}$ obtained in our analysis is $0.085\pm 0.003$ for sample sizes of 806 and 898. It might be argued that the $\sigma$ clipping is not a reliable procedure, as it could be manipulated to fit any desired model. However, this is not the case, as we follow a procedure with strict criteria governing the extent to which sources can be trimmed, as detailed earlier. We anticipate that this process will eventually reach a stage where further minimizing the scatter becomes practically impossible.

We clarify here that this procedure is employed to identify a subset of QSOs contributing to a standard set of features, where the intrinsic scatter is linked to the intrinsic scatter of these physical parameters. From a statistical perspective, we cannot further reduce this scatter. The remaining dispersion is attributed to the intrinsic physics of the QSOs. However, a comprehensive exploration of the correlations among these parameters goes beyond the scope of the current paper.
To select a `gold' sample, our initial step involves investigating the behaviour of the intrinsic dispersion $\delta_{int}$ within the luminosity-luminosity relationship for varying sample sizes. This analysis is depicted in Fig. \ref{fig:scatter}. The different sample sizes are determined using the $\sigma$-clipping procedure in the binning procedure described above; the corresponding values of interval selected for $\sigma$-clipping are illustrated in the colour bar of Fig. \ref{fig:scatter}.
The thickness of the rainbow line corresponds to 1 $\sigma$ error bars, while the colour of the line indicates the given $\sigma$ clipping. 
From Fig. \ref{fig:scatter}, it is evident that the trend is nonlinear. However, it is notable that the smallest $\delta_{int}$ values correspond to cases with fewer than 1300 sources and $\sigma$ clipping intervals between 0.7 (green) and 0.2 (red). Moreover, assuming the relation is linear and independent from redshift. In that case, we expect the slope and the scatter of this relation to be compatible for the whole sample after the trimming in each bin separately. This is fulfilled for a $\sigma$ clipping interval around 0.6.
We need to validate these intervals compared to the minimum achievable uncertainty when computing $\Omega_M$.
It is emphasized that our primary focus lies on checking the precision of the parameter $\Omega_M$. This parameter can be determined with the highest precision by fixing the value of $H_0$. 
The determination of the $H_0$ value is beyond the scope of this analysis.
To achieve the highest precision in determining $\Omega_M$, as previously mentioned, we must strike a balance between minimizing the intrinsic dispersion of the relation and maximizing the number of QSOs in the sample. A larger number of QSOs enhances the accuracy of our cosmological parameter estimation, provided we keep $\delta_{int}$ as small as possible. 
Hence, the criterion for defining the `gold' sample is to minimize the uncertainty on $\Omega_{M}$. Fig. \ref{fig:matter} illustrates that the smallest uncertainties on $\Omega_M$, plotted on the y-axis, correspond to $\sigma_{clipping} < 0.7$ values.

Additionally, Fig. \ref{fig:matter}, like Fig. \ref{fig:scatter}, illustrates that this trend is highly nonlinear. Similar to Fig. \ref{fig:scatter}, the color coding in the plot represents the interval chosen in the $\sigma$ clipping algorithm. 
Additionally, in Fig. \ref{fig:matter3}, we present the computed values of $\Omega_M$ along with their respective 1 $\sigma$ error bars. These results are presented as a function of the sample size obtained and the specific $\sigma$ clipping interval indicated by colour. It is worth noting that the precision of the matter density stabilizes at small intervals. For our final sample, we choose the biggest sample, corresponding to the $\sigma_{clipping}$ interval 0.6.
In Fig. \ref{fig:matter2}, the values of $\Omega_M$ are plotted as a function of the number of QSOs and are color-coded according to the highest probability of the $\Omega_M$ values. The most frequently occurring value is $\Omega_M=0.24 \pm 0.064$.

Furthermore, we establish that the `gold' sample should have similar properties to the initial sample. Thus, we performed a 2-sample Anderson-Darling test between the fluxes of both samples (`gold' and total). The distributions of redshift, $\log_{10}(L_X)$, and $\log_{10}(L_{UV})$ are compatible between the `gold' and the total sample at the levels of approximately 60\%, 16\%, and 73\%, respectively.

We stress that the $\sigma_{clipping}$ procedure was performed using the correction for evolution, as detailed in Appendix~\ref{AppendixB}. However, one can argue that, although the selection is independent of the particular choice of $\Omega_{M}$, it is based on the flat $\Lambda CDM$ model. We have previously demonstrated the correction for evolution in alternative cosmologies in~\cite{Dainotti2023alternative}. This is an interesting research endeavor, but it goes beyond the scope of the current paper.
In principle, there are various models to consider, including the open and closed universe, the wCDM model, and the numerous alternative cosmological models derived from the $f(R)$ theory of gravity. While one could apply the same procedure to all these models, it is reasonable to begin with the reference cosmological scenario, which is the $\Lambda$CDM model.


\section{Cosmological Fitting using the Golden Sample}\label{cosmology}
We have employed the final QSO sample to perform a fitting procedure using the method introduced by~\cite{Kelly2007}. This fitting is conducted within the framework of a flat standard $\Lambda$CDM model, where we set $H_0$ to three distinct values: 67.5, 70, and 73.5 $\; km\, s^{-1}\, Mpc^{-1}$. This analysis considers $\Omega_{M}$ as a free parameter with a wide uniform prior between 0 and 1.
Under these assumptions, we have also left the parameters of the Eq. \ref{lumevo}: $\gamma$', $\beta$', and its scatter $\delta$' free to vary, and we have imposed uniform priors as follows: $0 < \gamma' \leq 1$, $0 < \beta' < 20$, and $0 < \delta' < 1$. Hence, we have obtained the best-fit values of $\Omega_{M}$, $\gamma$', $\beta$', and $\sigma$' along with their associated 1 $\sigma$ uncertainties. Consequently, the results are presented in the left, middle, and right panels of Fig. \ref{fig:contour_gold}. In this fitting procedure, we have also considered the effects of the evolution in the redshift of QSO luminosities. To fit a cosmological model and explicitly show the dependence on $D_L$, we need to transform fluxes into luminosities according to Eq. \ref{eq:fluxes}. The redshift evolution of QSO luminosities has already been thoroughly examined, along with different possible approaches for correcting it in cosmological applications~\cite{Lenart:2022nip,DainottiQSO}. Here, we have applied the most general method for correcting for this evolution, known as `varying evolution', in which the correction and its uncertainties are functions of $\Omega_M$~~\cite{Lenart:2022nip}. This technique allows us to avoid the need for assuming a specific value of $\Omega_M$ to correct the luminosities for this evolutionary effect, thus entirely resolving the circularity problem.

As highlighted by~\cite{Dainotti2023alternative}, the QSO distribution of the RL relation does not exhibit a Gaussian distribution of residuals. Nevertheless, when comparing different samples that may follow different distributions, we utilize the Gaussian-based $\chi^2$ statistics as the basis for our likelihood.
\begin{equation}
  \chi^2= \left(\frac{log_{10}(L'_X)-{log_{10}(L'_{X,\, th})}}{s}\right)^2,
\end{equation}
where index $_{th}$ indicates the predicted value calculated as: $ log_{10}(L'_{X,th}) = \gamma' \times log_{10}(L'_{UV}) + \beta'$, and $s = \sqrt{\gamma^{\prime\, 2}\times \sigma^2_{log_{10}(L'_{UV})}+\sigma^2_{log_{10}(L'_{X})} + \delta_{int}^{\prime\, 2}}$
are the uncertainties. 
The data has no correlations between the individual data points. 
Here, $\sigma_{log_{10}(L'_{X/UV})}$ represents the uncertainty on the logarithm of the luminosity corrected for evolution in a particular band. With this definition, our likelihood is formulated as:

\begin{equation}
    log_{10}(PDF)= -\frac{1}{2}\sum_{i=1}^{N}log_{10}(s_{i}) -\frac{1}{2}\sum_{i=1}^{N}\chi_{i}^{2},
\end{equation}
where the index $_i$ represents the previously defined values for a particular source in the total sample of $N$ objects. This is the most common choice for the likelihood in the literature of the RL correlation. All our cosmological computations are conducted using the MCMC method. In this approach, all parameters of the correlation ($\gamma^{'}$, $\beta^{'}$, $\delta^{'}_{int}$) are allowed to vary along with $\Omega_{M}$. The contour plot for the sample obtained with an interval value of 0.6 of the $\sigma$ clipping, resulting in the smallest uncertainty on $\Omega_M = 0.240\pm 0.064$, is presented in the middle panel of Fig. \ref{fig:contour_gold}. Our study presents an examination of the dependency of the value of $\Omega_{M}$ and its associated uncertainty on the size of the sample obtained and the interval chosen. These findings are visually represented in Fig. \ref{fig:matter}.
Upon scrutiny of Table \ref{tab:results}, it is observed that all values of $\Omega_{M}$, as calculated in this study, are in concordance with those of the `gold' standard sample within a deviation of less than 0.3 $\sigma$. Furthermore, these values also agree with those obtained for the SNe Pantheon+ sample~\cite{2022ApJ...938..110B} ($\Omega_{M}=0.334\pm0.018$), within a deviation of less than 1.5 $\sigma$. 
To quantify the degree of compatibility, a z-score test is employed. This test is computed as: $\rm z=\frac{\Omega_{M,\, i}-\Omega_{M, comparison}}{\sqrt{\sigma_{\Omega_{M,\, i}}^2 + \sigma_{\Omega_{M, comparison}}^2}}$, where $\Omega_{M,\, i}$ and $\sigma_{\Omega_{M,\, i}}$ denote  our calculated value of $\Omega_M$ and its uncertainty as per Table \ref{tab:results}. The quantities with the subscript `comparison' pertain to the results obtained for the `gold' standard sample (second column to the right) and Pantheon+ (first column to the right).

\subsection{The independence of the results from $H_{0}$ initial values and a very mild evolution on $\Omega_M$}

\begin{figure}
    \includegraphics[width=0.32\textwidth]{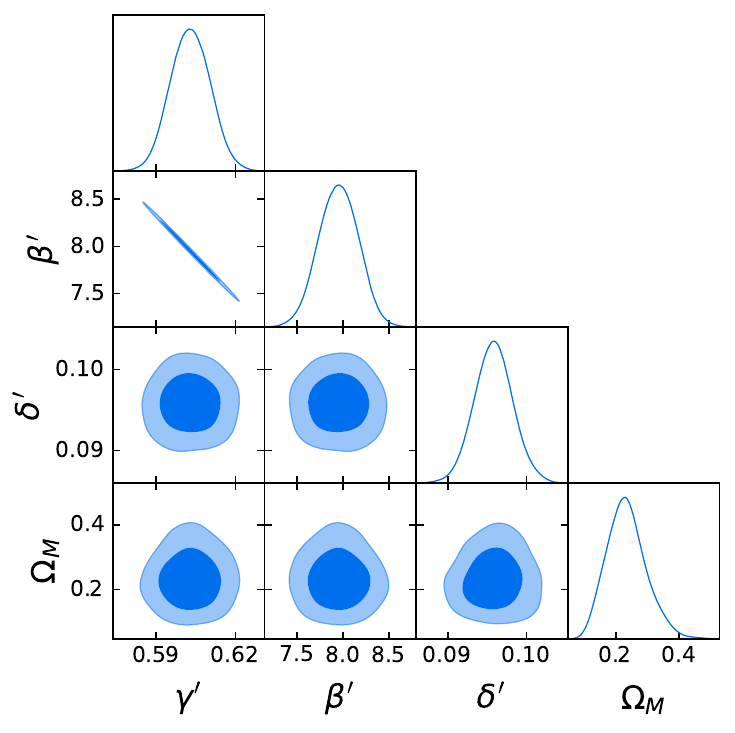}
    \includegraphics[width=0.32\textwidth]{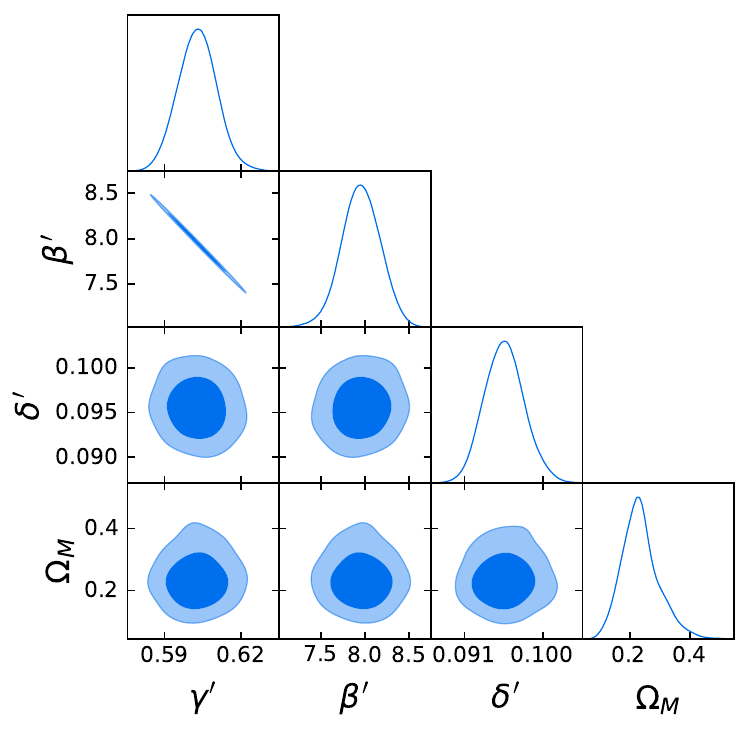}
    \includegraphics[width=0.32\textwidth]{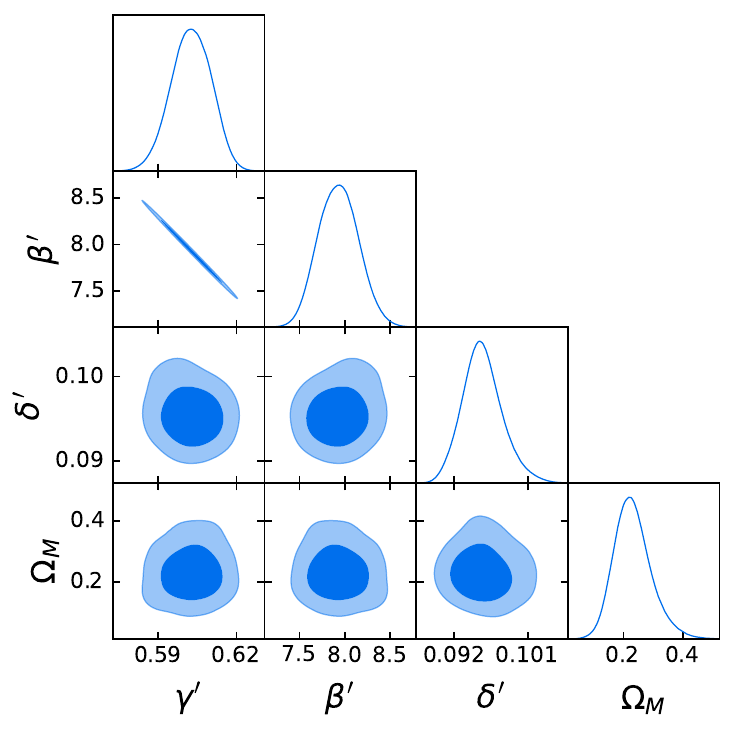}
    \caption{The contours for $\Omega_M$ for the `gold' sample, selected using a $\sigma$ clipping threshold of 0.6, are displayed in the figure. The results are depicted for three distinct values of the Hubble constant, with $H_0$ fixed at 67.5  $\rm km\, s^{-1}\, Mpc^{-1}$ (on the left), $H_0= $70 $\rm km\, s^{-1}\, Mpc^{-1}$ (in the middle), and $H_0 = $73.5 $\rm km\, s^{-1}\, Mpc^{-1}$  (on the right panel).}
    \label{fig:contour_gold}
\end{figure}

We demonstrate that our results remain independent of the initial parameters of $H_0$. We conducted the entire analysis using three different values of the Hubble constant: $H_0=70$, $73.5$ and $67.5\, \rm km\, s^{-1}\, Mpc^{-1}$. We choose these specific values of $H_0$ as the first is the fiducial value in~\cite{scolnic2018}, the second is derived from measurements involving a set of 42 Cepheids \citep{Riess2019ApJ...876...85R}, and the third is obtained from Planck data of the CMB \citep{Planck:2018vyg}, respectively. The results for $\Omega_M$ obtained for each of these values are presented in Table \ref{tab:results}, showing no significant differences between these cases.
Indeed, they all agree within 1 sigma.
In Fig. \ref{fig:contour_gold}, we present the posterior results of the fitting for the `gold' sample considering values of $H_0=70$, $73.5$ and $67.5\, \rm km\, s^{-1}\, Mpc^{-1}$.
To further guarantee that this independence is not limited solely to the `gold' sample, we conducted the same analysis on all other subsamples and always obtained consistent results. For the `gold' sample (interval of the $\sigma$- clipping 0.6), we derived the following values: $\Omega_{M, H_0=67.5} = 0.236\pm 0.063$, $\Omega_{M, H_0=70} = 0.240\pm 0.064$, and $\Omega_{M, H_0=73.5} = 0.231\pm 0.062$. It is important to mention that the EP method remains invariant under linear data transformations. Thus, the chosen value of $H_0$ does not impact the results when this correction is applied. The change in $H_0$ values rescales the luminosity uniformly across all redshifts, representing a linear transformation in this context. It's worth noting that to compute the $\delta_{D_{L}}$ parameter for a specific bin, a fixed value of $\Omega_M$ is required. Consequently, our binning division depends on this value. However, when we varied this parameter across a range of values, no significant differences were observed in the results.

Regarding the consideration of the evolution in $\Omega_M$, we can look at the QSO Hubble diagram, which either suggests an evolution of cosmological parameters, such as $\Omega_M$, either exclusively at higher redshifts, as claimed in \cite{perlmutter1999}, or throughout the sample, as claimed in \cite{Planck:2018vyg}. The work by \cite{Risaliti_Athena} highlights a decline in the distance modulus, and attributing a higher value to $\Omega_M$ presupposes the $\Lambda CDM$ model. While the reduction in this decline compared to Planck is notable due to the luminosity correction, there appears to be some residual evolution in $\Omega_M$, which will be further discussed in section \ref{cosmologyf(R)}. 
\section{Interpretative paradigm via modified gravity}\label{MG}

The data analysis conducted in the preceding sections implies the potential for residual evolution in the parameter $\Omega_{M}$ when examining sources more distant than SNe Ia. This perspective could also have noteworthy implications for resolving the $\sigma_8$-tension \cite{DES:2021wwk, KiDS:2020suj, Philcox:2021kcw, Kazantzidis:2018rnb, DiValentino:2020vvd}. Hence, within this context, we delve into a viable scenario that theoretically accommodates a varying $\Omega_{M}$ with redshift.
Given the slight variation in the mean value of $\Omega_M$, we explore potential causes. In this section, we examine the feasibility of a theoretical model that could explain an effective variation of $\Omega_M$ with redshift.
The idea is that, as discussed in ~\cite{Dainotti:2021pqg, Dainotti:2022bzg,Schiavone:2022wvq,Montani2023MNRAS.tmpL.155M} regarding the $H_0$ tension. It posits that the parameter $\Omega_M$ might be subject to an underlying physical phenomenon, contributing to its apparent variation with redshift.

Therefore, within this framework, we utilize a modified gravity approach, specifically a metric $f(R)$-gravity model in the Jordan frame, as detailed in references~\cite{Sotiriou-Faraoni:2010, Nojiri:2010wj}. When applied in a flat Robertson-Walker geometry with a matter component, the two fundamental equations of the proposed framework can be expressed as follows:

\begin{equation}
	H^2 = \frac{1}{\phi}\left( 
	\frac{\kappa^2}{3}\rho_m - H\dot{\phi} + \frac{V(\phi )}{6}\right),
	\label{mdv1}
\end{equation}
and 
\begin{equation}
	R \equiv \frac{dV}{d\phi} = 6\dot{H} + 12H^2. 
	\label{mdv2}
\end{equation}
Above, $\kappa^2 = 8 \pi G$ where $G$ is the gravitational constant and $c=1$. Here, $H\equiv \dot{a}/a$ (where the dot indicates differentiation with respect to the synchronous time) represents the Hubble parameter, and $\rho_m$ stands for the matter-energy density. 
Additionally, the second equality in Equation (\ref{mdv2}), along with Equation (\ref{mdv1}), results in the Klein-Gordon-like equation for the scalar field in an isotropic Universe, expressed as:

\begin{equation}
3\ddot{\phi} + 9H\dot{\phi} 
- 2V(\phi ) + \phi \frac{dV}{d\phi} = \kappa^2 \rho_m
\, . 
\label{13bis}
\end{equation}

Nonetheless, for our purposes, Equation (\ref{mdv2}) is more appropriate than the latter.
This formulation is entirely equivalent to the one in the review by \cite{DeFelice:2010aj}; specifically, we inverted the relations $\phi = df/dR \equiv F$ and $V = RF - f$, where $R$ represents the Ricci curvature scalar.
In terms of the cosmic scale factor $a(t)$ (with its present-day value conventionally set to unity), the expression for $\rho_m$ is as follows:

\begin{equation}
	\rho_m = \frac{\rho_m^0}{a^3} \equiv \rho_m^0\left( 1 + z\right)^3,
	\label{mdv3}
\end{equation}
where $\rho_m^0$ represents the present value of the matter-energy density.

Here, the scalar field $\phi$ exhibits a non-minimal coupling to standard gravity, specifically to the dynamics of the scale factor. In the Jordan frame, the potential term $V(\phi)$ corresponds to the specific form of the function $f(R)$. In particular, we have the following relation between these two quantities:

\begin{equation}
	f(R/\phi )) = \phi \frac{dV}{d\phi} - V(\phi )
	\, , 
	\label{mdv4}
\end{equation}
where the expression for $R(\phi)$ is determined by reversing the definition provided in Equation (\ref{mdv2}).

Clearly, Equation (\ref{mdv1}) corresponds to the generalised Friedmann equation, whereas Equation (\ref{mdv2}) is obtained by varying the action with respect to the scalar field $\phi$. 

We now aim to achieve two simultaneous and complementary effects: i) we intend to generate the vacuum energy density of the present Universe through the potential term of the scalar field $\phi$; ii) we need to construct a dynamic scenario in which the matter-energy density and the vacuum energy density exhibit different behaviours with respect to the redshift. In particular, the former must be suppressed in comparison to the latter.

These two features are easily obtained if we require that the potential term be expressed as follows:

\begin{equation}
	V (\phi ) = 2 \kappa^2 \rho_{\Lambda}\phi + G (\phi )
	\, , 
	\label{mdv5}
\end{equation}
where $\rho_{\Lambda}$ denotes the Universe vacuum energy density, while 
$G(\phi )$ is a generic correction. 
Therefore, Equation (\ref{mdv1}) can be rewritten as:

\begin{equation}
	H^2 = \frac{\kappa^2}{3}\left( \frac{\rho_m}{\phi} + \rho_{\Lambda}\right) 
	+ \frac{1}{\phi}\left(-H\dot{\phi} + \frac{G(\phi )}{6}\right).
	\label{mdv6}
\end{equation}

Considering the result obtained in the data analysis above, it is natural to split the equation above into the following two:

\begin{equation}
	H^2 = H_0^2 \left( \frac{\Omega_{M}}{\phi}(1 + z)^3 + \Omega_{\Lambda}\right)
	\, 
	\label{mdv7}
\end{equation}
and
\begin{equation}
	6H \dot{\phi} = G(\phi )
	\, .
	\label{mdv8}
\end{equation}

We observe that this position corresponds to setting the second term in the right-hand-side of Equation (\ref{mdv5}) equal to zero, which is, in principle, a special choice. However, we stress that the field equations are not linear, and thus, they do not admit a unique solution. We are searching for a specific behaviour that can induce an effective evolution of $\Omega_{M}$, while remaining as close as possible to a $\Lambda CDM$-Hubble parameter.
The position mentioned above precisely fulfills this requirement, and it was successfully employed in the recent previous analysis \cite{Montani:2023xpd}, addressing a similar question regarding the Hubble tension.
The value of constraining the space of possible solutions in this manner lies in the flexibility of the obtained model when compared with the data, as only the evolution of the scalar field $\phi$ contributes to the luminosity distance. Furthermore, the idea of introducing a new equation allows us to follow a new paradigm in which the form of the $f(R)$ and its viability are determined a posteriori. Our goal is to determine which type of modified gravity is capable of generating the desired effect rather than testing different $f(R)$ models (thus assigning a priori the potential term $4V(\phi )$).
In Equation (\ref{mdv7}), we introduced the Hubble constant $H_0$ and the two critical parameters $\Omega_{M}$ and $\Omega_{\Lambda}$ for the matter and vacuum energy, respectively (here, we assume that $\Omega_M\simeq 0.3$ and $\Omega_{\Lambda}\simeq 0.7$). 
It is important to emphasise that, in order to remain consistent with experimental constraints on the validity of General Relativity, the current value of $\phi$ should be very close to unity (with deviations on the order of $10^{-7}$). Therefore, at present and for sufficiently small redshift values, Equation (\ref{mdv7}) aligns with the predictions of the $\Lambda CDM$ model.

Now, by introducing the time variable $x\equiv \ln (1 + z)$ and noting that the relation $d(...)(dt) = -Hd(...)/dx$ holds, Eqs (\ref{mdv7}), (\ref{mdv2}), and (\ref{mdv8}) can be rewritten as follows:

\begin{eqnarray}
	H^2 = H_0^2
	\left( \frac{\Omega_{M}}{\phi}e^{3x} + \Omega_{\Lambda}\right)
	\, , \label{mdv9}\\
	2\kappa^2 \rho_{\Lambda} + \frac{dG}{d\phi} = 6H\left(2H - \frac{dH}{dx}\right)
	\, , \label{mdv10}\\
	6H^2\frac{d\phi}{dx} = - G(\phi )
	\, , \label{mdv11}
\end{eqnarray}
respectively. These three equations enable us to determine the evolution of the three unknowns: $H(x)$, $\phi(x)$, and $G(x)$ respectively. 
We have solved these equations with the following steps: a) diving all the equations by $H_0^2$, b) introducing  the two dimensionless variables $u=(H/H_0)^2$ and $w(\phi(x))=G(\phi)/H_0^2$, c) using the equality $2\kappa^2 \rho_{\Lambda}/H_0^2=6 \Omega_\Lambda$,  and d) multiplying the equation 20 by $\frac{d\phi(x)}{dx}$. Thus, we can re-write the system of equations (19)-(21) in the following dimensionless form:

\begin{eqnarray}
	u(x) = 
	\left( \frac{\Omega_{M}}{\phi}e^{3x} + \Omega_{\Lambda}\right)
	\, , \label{mdv12}\\
	 \frac{dw}{dx} = (12u(x) -3 \frac{du(x)}{dx} -6{\Omega_\Lambda})\frac{d\phi(x)}{dx} 
	\, , \label{mdv13}\\
6u(x)\frac{d\phi}{dx} = 	- w(\phi(x)) 
	\, , \label{mdv14}
\end{eqnarray}

From the last two quantities, we can determine the potential term $G(\phi)$ and, thus, establish the form of the considered modified gravity Lagrangian.
We emphasize that if $G(\phi) < 0$ within a certain $\phi$-interval, then according to Equation (\ref{mdv11}), it guarantees that $d\phi /dx > 0$. Consequently, the function $\phi(x)$ increases with the redshift from unity to a larger specified value, and the contribution of $\Omega_{M}$ is accordingly diminished compared to a standard $\Lambda CDM$ model.
Clearly, the detailed dynamics of the scalar field play a crucial role in accounting for potential variations in the importance of the parameter $\Omega_{M}$ throughout the evolution of the Universe, as observed through various astrophysical and cosmological sources. 
The proposed concept of a varying parameter $\Omega_{M}(z)$ should ideally be integrated into a comprehensive framework alongside the one proposed in ~\cite{Montani:2023xpd} to address the so-called Hubble tension. Establishing a unified scheme of fundamental physics to deal with cosmological parameters is essential for creating a robust tool for testing data.

In addition, one can reduce the system of the above 3 equations to a single non-linear second-order differential equation for $\phi(x)$. Differentiating equations (\ref{mdv12}) and (\ref{mdv14}) with respect to x and substituting into equation (\ref{mdv13}), one obtains:

\begin{equation}
    2 \left(\Omega _M-1\right) \phi (x)^2
   \left(\phi '(x)+\phi ''(x)\right)+e^{3 x}
   \Omega _M \left(\phi '(x)^2-\phi (x)
   \left(7 \phi '(x)+2 \phi
   ''(x)\right)\right)=0
   \label{diff_phi}
\end{equation}

\begin{figure}
  \includegraphics[width=0.9\textwidth]{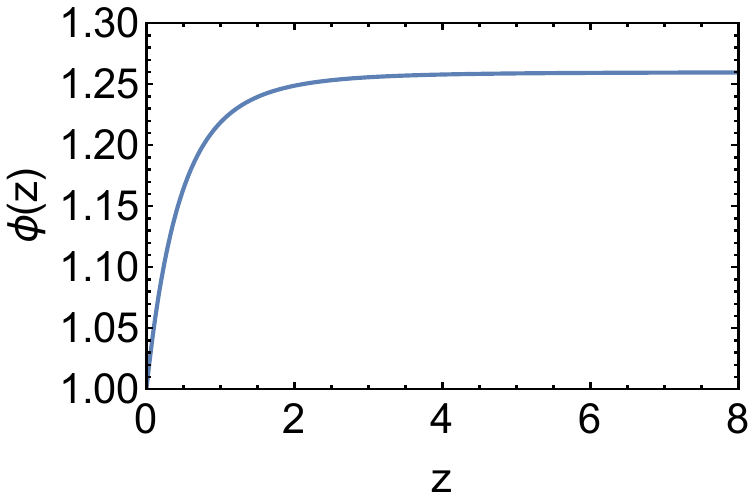}
  \caption{The solution for $\phi(z)$ by assuming $\Omega_M=0.30$ and $w_0=-3.4$ in Eq. \ref{mdv13}.}
  \label{phi(z)}
\end{figure}

We have solved the equations above by imposing the following initial conditions: $\phi(x_0)=1$, and $\phi'(0)=\frac{-w(0)}{6u(0)}=\frac{-3.4}{6}$.
We here plot the $\phi(z)$ vs z solution in Fig. \ref{phi(z)}. We see that the scalar field slightly increases, reaching a plateau for
$z \propto 2$. As a consequence, the value of the effective matter critical parameter, defined as $\Omega_M/\phi(x)$, slightly decreases from the mean value obtained from the initial imposed value $\Omega_M=0.30$, up to the mean value emerging from our QSO analysis. 

\section{Cosmology with the modified gravity scenario}\label{cosmologyf(R)}
\begin{figure}
  \includegraphics[width=0.9\textwidth]{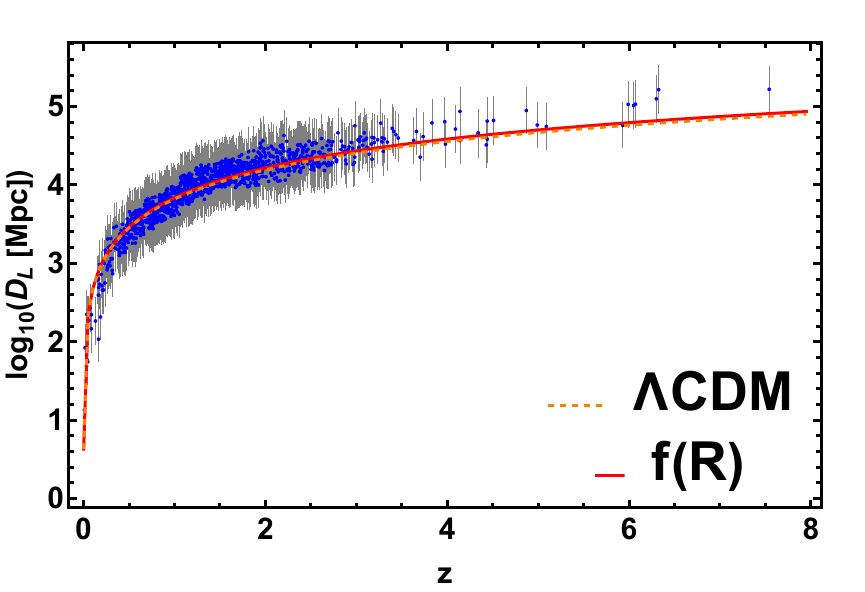}
  \caption{The distance luminosity in the $\Lambda$CDM and induced by the f(R) modified theory of gravity for $\Omega_M$=0.3 are shown in dashed orange and continuous red, respectively, with the total QSO sample in grey and the `gold' sample in blue.}
  \label{DL-LambdaCDMvsf(R)}
\end{figure}

\begin{figure}
  \includegraphics[width=0.85\textwidth]{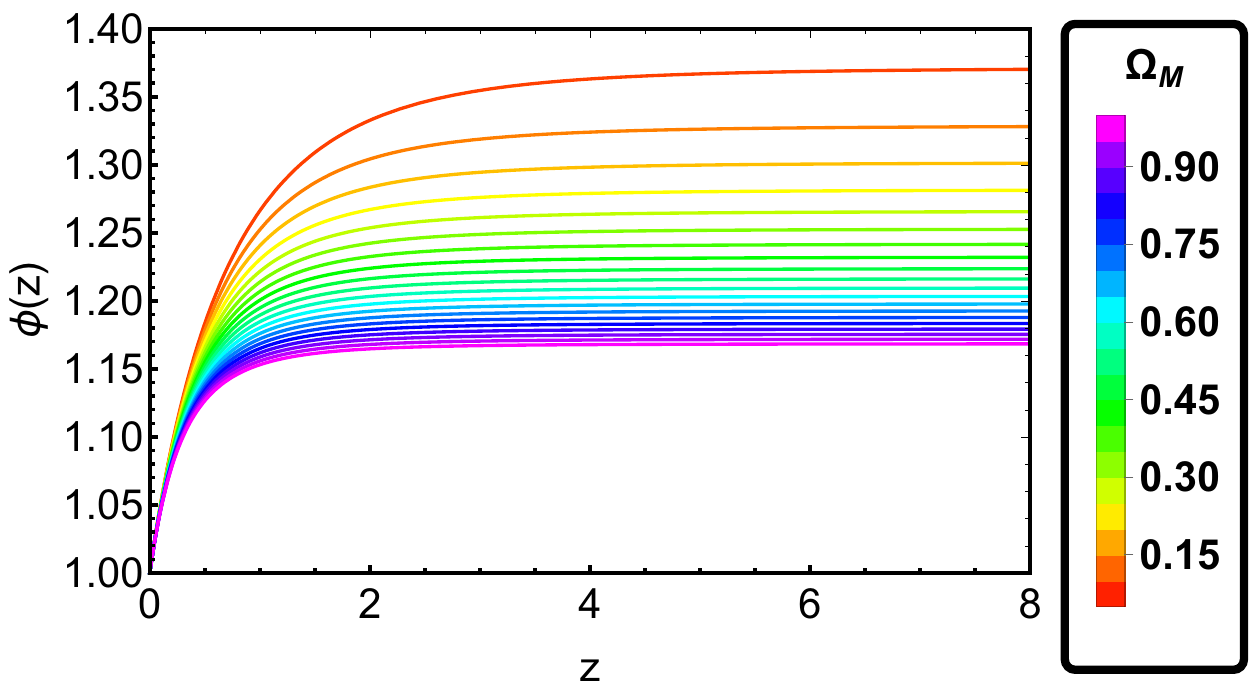}
   \includegraphics[width=0.85\textwidth]{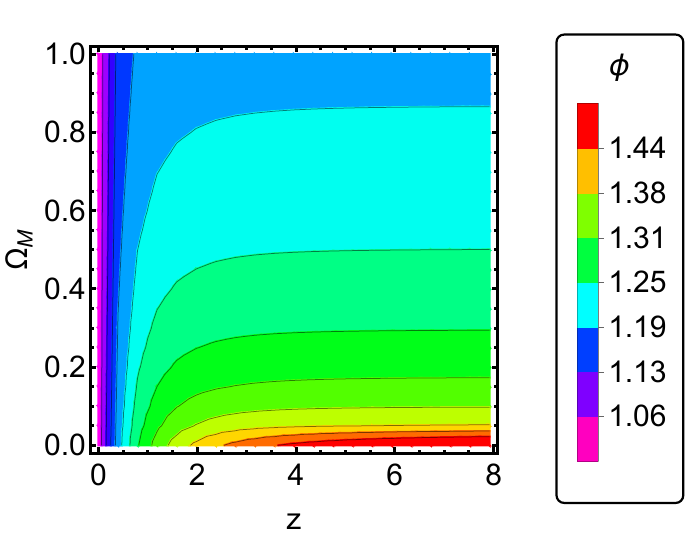}
  \caption{Upper panel: the solution for $\phi(z)$ vs. z as a function of values of $\Omega_M$ ranging from 0.05 to 1 colour-coded in the colour bar in the right panel and assuming $w_0=-3.4$ in Eq. \ref{mdv13}. Lower panel: The redshift as a function of $\Omega_M$ colour-coded according to the values of the $\phi(z)$.}
  \label{phivszvsOmegaM}
\end{figure}

The modification in the critical matter parameter is reflected in a corresponding change in the luminosity distance, which now takes the following form: 

\begin{equation}
D_{L} = (1+z) \frac{c}{H_{0}} \int_{0}^{z} \frac{dz'}{\sqrt{(\Omega_{M}/\phi (z^{'})) (1+z')^3 + (1-\Omega_{M})}},
\end{equation} 
where $\phi(z)$ is the function plotted in Fig \ref{phi(z)} (for fixed $\Omega_M=0.3$). 
To check the departure of the luminosity distance between the $\Lambda CDM$ and the one induced by the f(R) theory of gravity, we have plotted the logarithm of $D_L$ with both models in Fig.~\ref{DL-LambdaCDMvsf(R)} considering a fixed fiducial value of $\Omega_M=0.3$. We can see that the two distance luminosities are the same up to $z=1$, and then they start to depart at a later time (z). 
We emphasise that since each solution of the $\phi(z)$ is a function of $\Omega_M$, thus to study the trend of $\phi(z,\Omega_M)$ we have created a grid of 20 $\Omega_M$ values from 0.05 to 1 to obtain a series of $\phi(z)$ functions shown in the upper panel of Fig.~\ref{phivszvsOmegaM}.
To ensure uniformity in the behaviour of this function $\phi(z)$, we choose the values of $w_0=-3.4$ used in the previous section.
In the lower panel of Fig. \ref{phivszvsOmegaM} we show this trend in a different way, namely $\Omega_M$ as function of z color-coded with the $\phi(z)$. Thus, the presence of the varying scalar field $\phi(z)$ induces a corresponding scaling of $\Omega_M$, which can be determined with different values for sources at different redshifts.
However, to estimate the extent to which the variation in the cosmological setting impacts the determination of  $\Omega_M$, one would need to repeat the same analysis as above but using the new distance luminosity. Moreover, it is crucial to recompute the evolutionary parameters for the QSOs `gold' sample considering this new distance luminosity within a modified theory of gravity. 
However, we would like to stress that if we aim for a complete procedure for the cosmological analysis we would need to change the values of $w_0$ together with redshift and $\Omega_M$. This would require a thorough investigation in four dimensions of the behaviour of $\phi(z)$ as a function of $w_0$, $\Omega_M$, and z, which, in turn, would enlarge the uncertainties on $\Omega_M$. 

\section{Summary and Conclusions}\label{conclusions}

To identify a `gold' QSO sample with the most precise relation for efficient cosmological applications, we utilize the most recent and comprehensive QSO dataset dedicated to cosmological investigations ~\cite{2020A&A...642A.150L}. This dataset comprises 2,421 sources, covering a redshift range from $z = 0.009$ to $z = 7.54$. Importantly, our approach differs from previous studies as we utilize the complete sample across all redshifts. Furthermore, we implement redshift evolution corrections for cosmological applications following 
the methodology outlined in
~\cite{Lenart:2022nip,Dainotti:2022rfz}. This strategy allows us to retain sources at lower redshifts (i.e., $z \leq 0.7$), typically excluded in most prior analyses.
Our approach for identifying this `gold' sample is versatile and applicable to other probes (e.g., GRBs) and larger datasets. Additionally, this method is entirely model-independent, thereby avoiding the issue of circular reasoning. Specifically, we employ the $\sigma$-clipping technique on the linear relation between the logarithms of QSO fluxes in UV and X-ray. Notably, this approach is not reliant on a specific cosmological model. For the first time in the literature, we utilized this innovative procedure, characterized by independence from cosmological models and a robust statistical foundation.

As mentioned, our method relies on the so-called $\sigma$-clipping technique, applied in small bins. This technique enables us to minimize the intrinsic dispersion of the $log_{10}(F_{X}) - log_{10}(F_{UV})$ relation by eliminating sources with a vertical distance from the best-fit relation exceeding a defined threshold value. 
We acknowledge that sources scattered around this relation pose a challenge in identifying the most suitable sample with minimal intrinsic dispersion. To overcome this issue, we have applied the well-established statistical technique, the Theil-Sen method. This robust method is used to estimate the parameters of a model, specifically the $log_{10}(F_{X}) - log_{10}(F_{UV})$ relation within bins, serving as a model-independent approximation of the $log_{10}(L_{X}) - log_{10}(L_{UV})$ relation. 
Utilizing this robust method allows us to identify outliers and give them less weight when estimating the best-fit parameters of the model. With this adopted technique, we have identified a sample of QSOs that acts as a standard candle with a reduced dispersion in fluxes, $\delta = 0.096\pm 0.003$, representing a 56\% reduction compared to the dispersion of the original sample ($\delta_{int} = 0.22$). It is important to note that this sample has been constructed based on the best-fit linear relation in the $log_{10}(F_{X}) - log_{10}(F_{UV})$ space in each bin separately~\cite{2020A&A...642A.150L}. After applying the $\sigma$-clipping method, performed in bins of redshifts small enough to ensure that the differences in luminosity distance, $\Delta_DL$, are negligible compared to the dispersion of the relation, we have identified a sample comprising 1253 QSOs. This sample is smaller in size compared to the original dataset (with $\approx 48\%$ of sources discarded), yet it retains the same essential characteristics as the parent sample. The distributions of $log_{10}(L_X)$, $log_{10}(L_{UV})$, and $z$ are found to be compatible in the two samples at a significance level greater than 0.05. 
In each bin, we applied the Anderson-Darling two-sample test to verify that the two distributions in fluxes are drawn from the same parent population as the total initial sample. The reduction in the sample size is understandably substantial, with the 1048 Pantheon SNe Ia slimmed down from an original sample of 3473 events, constituting a reduction in size of 70\% of the starting dataset~\cite{scolnic2018}. Importantly, adopting this technique in narrow redshift bins not only facilitates thorough investigations but may also completely eliminate any dependency on distance.

Our main goal is to define a sample that can effectively constrain cosmological parameters, such as $\Omega_M$. Therefore, we aim to achieve not only the smallest dispersion of the $log_{10}(L_{X}) - log_{10}(L_{UV})$ relation but also to maintain a statistically sufficient number of sources in each redshift bin. This balance ensures that the fitting process remains viable and statistically robust. Therefore, we have found a compromise between minimizing dispersion and maintaining the largest feasible sample size, even though these two goals are somewhat contradictory. 
The optimal number of sources found using this method is 1253, determined by applying a $\sigma$-clipping threshold of $0.6$ in each redshift bin.
We emphasize that our procedure also necessitates meeting additional criteria: maintaining a minimum of three sources per bin, passing the Anderson-Darling two-sample tests in each bin, and ensuring that the luminosity distance variation is negligible compared to the dispersion of the flux relation in each bin.

Considering the balance between the number of sources and the reduction in intrinsic dispersion, as previously discussed, the selected sample of 1,253 QSOs has an intrinsic dispersion, $\delta_{int} = 0.096$, and still extends up to the maximum redshift of z = 7.54 of the original sample. These 1,253 QSOs, selected as described from the `gold' sample, are shown in Fig. \ref{fig:goldensample}, along with the corresponding best fit of the $F_{X} - F_{UV}$ relation. To test and validate our approach, we have shown that this `gold' sample still adheres to the intrinsic RL relation after adjusting for selection effects and luminosity evolution with redshift following the methodology outlined in~\cite{DainottiQSO, 2011ApJ...743..104S}. 

As outlined in Section \ref{EPcorrection}, the `gold' sample has produced an intrinsic normalization comparable to the corresponding $L'{X} - L'{UV}$ relation for the original sample~\cite{DainottiQSO}. The values of the slope are consistent within <1 $\sigma$, and the normalization is consistent within 1.5 $\sigma$.

Subsequently, we applied this `gold' sample for cosmological purposes to estimate $\Omega_M$ by fitting a flat $\Lambda CDM$ model. We kept both $\Omega_M$ and the RL relation parameters, $\gamma'$, $\beta'$, and $\delta'_{int}$, free, while fixing $H_0 =70 \rm km\, s^{-1}\, Mpc^{-1}$. Furthermore, we implemented corrections for the redshift evolution of luminosities ($L_{X}$, $L_{UV}$) by applying corrections as functions of $\Omega_M$ (i.e., $g(z, \Omega_M)$ and $h(z, \Omega_M)$). This correction method has been successfully employed in previous cosmological analyses to avoid circularity issues ~\cite{Lenart:2022nip,Dainotti2023alternative,Bargiacchi2023MNRAS.521.3909B}. By incorporating the evolutionary parameters and without assuming any specific cosmological model, we achieved a significant reduction in the uncertainty associated with $\Omega_M$.
We derived $\Omega_M = 0.240 \pm 0.064$ using the `gold' sample, as illustrated in Fig. \ref{fig:contour_gold}. This value aligns with the current estimate of $\Omega_M = 0.334 \pm 0.018$ obtained with SNe Ia~\cite{2022ApJ...938..110B}, differing by less than $1.5\sigma$.
This smaller mean value of $\Omega_M$ finds theoretical support in Section \ref{MG}. Moreover, the precision of the $\Omega_M$ uncertainty, measured at $0.064$, is comparable to the precision of $0.10$ achieved by Type Ia Supernovae in 2011.

In conclusion, our results signify a breakthrough, showcasing the capability of QSOs as reliable standard candles for the precise determination of cosmological parameters, including $\Omega_M$, achieving a level of precision comparable to SNe Ia but at redshifts up to 7.5.
This marks a new era for QSOs as reliable standard candles, opening avenues for the QSO community to explore and comprehend the distinctions between this `gold' sample and the overall dataset. Our analysis highlights the significance of future missions, like the Athena mission~\cite{Athena2013}, capable of observing QSOs with significantly enhanced precision.

\begin{acknowledgments}

The authors are particularly grateful to Giada Bargiacchi for her helpful discussion and contribution to the initial software development. EDV is supported by a Royal Society Dorothy Hodgkin Research Fellowship.  This article is based upon work from COST Action CA21136, Addressing observational tensions in cosmology with systematics and fundamental physics (CosmoVerse) supported by COST (European Cooperation in Science and Technology).
\end{acknowledgments}

\section{Appendix}

\subsection{Correction for evolution}
\label{EPcorrection}

Here, we provide a short explanation of the Efron \& Petrosian (EP) methods in our analysis. We applied the same algorithm used by~\cite{DainottiQSO,Lenart:2022nip}. The EP method is central to determining the correction parameter applied to our data. We apply it as outlined below. We assume that the correct distribution of luminosity can be obtained by dividing the observed quantities by a given function of the redshift: 

\begin{equation}
    L_{corrected} = \frac{L_{observed}}{g(z)},
\end{equation}

where $L_{observed}$ represents the observed luminosity distribution within a specific band, $L_{corrected}$ is the true distribution (the distribution which would be observed in the absence of selection bias and redshift evolution), and $g(z)=(1+z)^{k_{L}}$ is the correction as a function of redshift. The EP method uses a non-parametric test to determine the parameter value $k_L$. 
It searches for the value of $k_L$ that, upon applying the correction, gives the distribution of luminosity uncorrelated with redshift. For more in-depth details, see this application for GRBs in ~\cite{Efron1992ApJ...399..345E,Dainotti2013a,Dainotti2013b,dainotti17a,
Levine2022ApJ...925...15L}. In this work, our primary focus is on cosmological computations. Thus, following ~\cite{Lenart:2022nip, DainottiQSO} we consider $k_L$ as a function of the cosmological parameter, $\Omega_M$. This plays a crucial role in applying it in the Monte Carlo Markov Chain (MCMC) fitting process while avoiding any prior assumptions of a fixed value of $\Omega_M$. We show how $k_{L_{X}}$ and $k_{L_{UV}}$ changes with different values of $\Omega_M$ at Fig. \ref{fig:k} in the Appendix \ref{AppendixB}. 
A detailed technical description and the code for this method are available in the Wolfram Mathematica Notebook Archive~\cite{EP_notebook}. Thus, the equation describing the correlation of luminosities, corrected for selection bias and redshift evolution, as a function of $\Omega_M$ is expressed as:

\begin{equation}
  \log_{10}(L_X) - k_{L_{X}}(\Omega_M)\times \log_{10}(1+z) = \gamma'\times (\log_{10}(L_{UV}) - k_{L_{UV}}(\Omega_M)\times \log_{10}(1+z))+\beta'.
  \label{lumevo}
\end{equation}

\begin{figure}[ht!]
\includegraphics[width=0.49\textwidth]{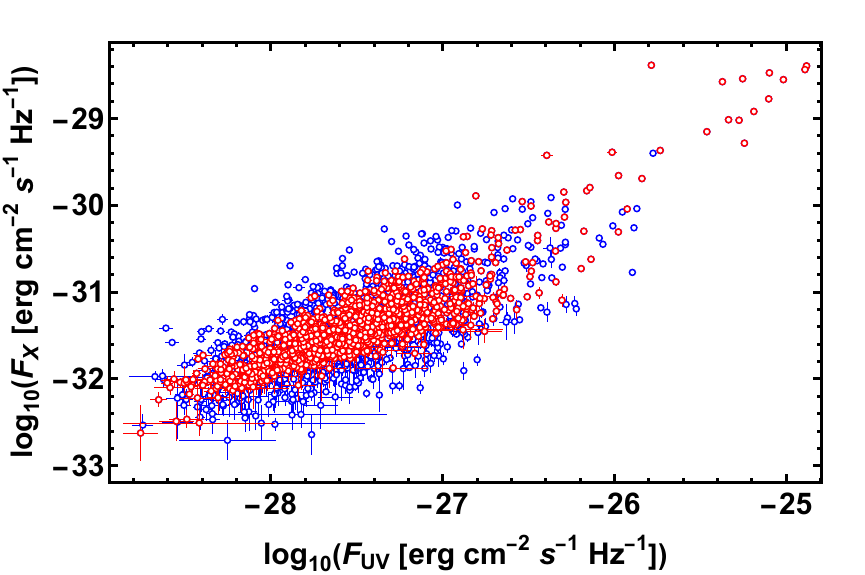}
\includegraphics[width=0.49\textwidth]{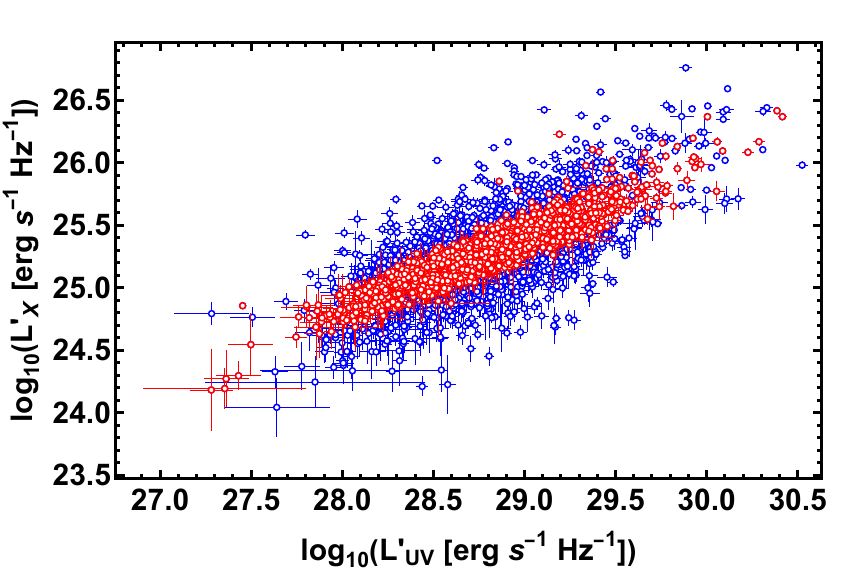}
\caption{The left panel displays the $log_{10}(F_{X}) - log_{10}(F_{UV})$ relation, while the right panel shows the $log_{10}(L^{'}_{X}) - log_{10}(L^{'}_{UV})$ relation, both corrected using the Efron \& Petrosian method. The `gold' sample is depicted in red, and the entire sample is in blue.}
  \label{fig:goldensample}
\end{figure}

Further, in the text we use the following notation: $log_{10}(L'_{X})=log_{10}(L_{X})-k_{L_{X}}\times log_{10}(1+z)$, $log_{10}(L'_{UV})=log_{10}(L_{UV})-k_{L_{UV}}\times log_{10}(1+z)$. We have identified the `gold' sample in the main text, which is represented in red, both in the left panel of Fig. \ref{fig:goldensample} for the flux-flux distribution and in the right panel of Fig. \ref{fig:goldensample} for the luminosity-luminosity distribution. This latter distribution has been corrected using the (EP) ~\cite{Efron1992ApJ...399..345E} method according to Eq. \ref{lumevo} with $\Omega_{M}=0.3$. In our fittings performed with the MCMC method, we obtained for the defined above `gold' sample: $\gamma' = 0.603\pm 0.008$, $\beta' = 7.93\pm 0.23$, and $\delta'_{int} = 0.095\pm 0.002$. It is essential to notice that these results are in excellent agreement with those obtained by~\cite{Lenart:2022nip} within 1.5 $\sigma$ ($\gamma'= 0.591\pm 0.013$, $\beta'=8.278\pm 0.0362$), but with much smaller scatter, which for the whole sample is $\delta'_{int} = 0.231\pm 0.004$. 

We note that we have also tried the binning procedure with fluxes corrected for evolution ($log_{10}(F_X)-k_{L_{X}}\times log_{10}(1+z)$, and similarly with corrected UV fluxes). We did not observe statistical differences with the analysis presented here. Thus, we opted only with the fitting-in bins not corrected for evolution because it is a more straightforward approach. 

It is also important to highlight, as evident from Fig. \ref{fig:goldensample}, that the sources predominantly cluster in the middle of the distribution. No distinct trend is observed in these sources concerning their uncertainties, either in luminosities or fluxes. For a comprehensive analysis of the QSO 'golden' sample, an investigation into all its characteristics compared to the entire parent population would be required. However, such an in-depth study falls outside the scope of this paper.

\subsection{$\Omega_M$ independent correction for evolution}\label{AppendixB}
To correct for redshift evolution and selection bias in luminosity, it is necessary to calculate luminosity using observed flux and redshift. The luminosity requires an assumption of the values of cosmological parameters. Thus, the evolutionary function $g(z)=(1+z)^k$ in formula $L_{corrected} = L_{observed}/g(z)$ depends on cosmological parameters chosen a priori. This dependence takes the form $k=k(\Omega_M)$. The numerical behavior of the function $k(\Omega_M)$ can be determined by computing the $k$ parameter across a range of $\Omega_M$ values and approximated by a cubic spline interpolation as presented by \cite{Dainotti2023MNRAS.518.2201D, Dainotti:2022rfz, Lenart:2022nip}. This approach yields the corrected luminosity expressed as a function of $\Omega_M$:

\begin{equation}
    L_{corrected}(\Omega_M)=L_{observed}(\Omega_M)/(1+z)^{k(\Omega_M)}.
\end{equation}

\begin{figure}
    \includegraphics[width=0.45\textwidth]{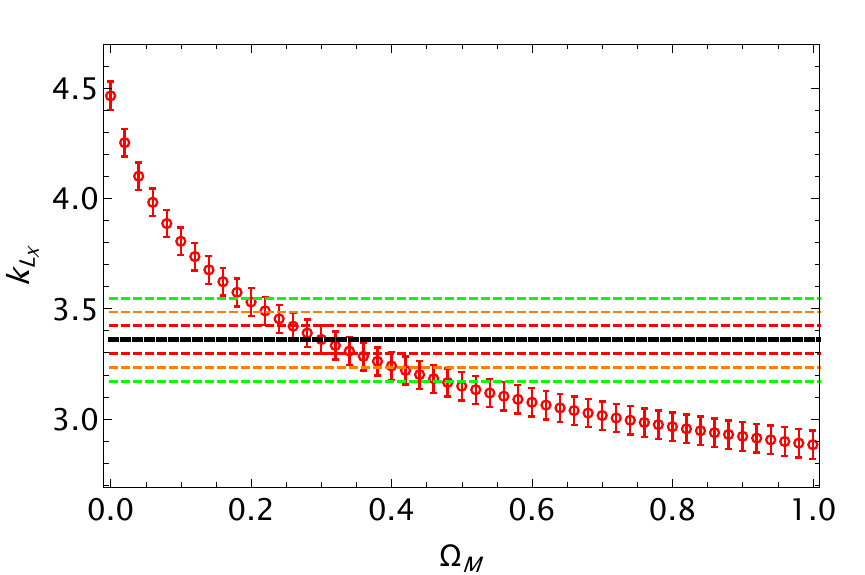}
    \includegraphics[width=0.45\textwidth]{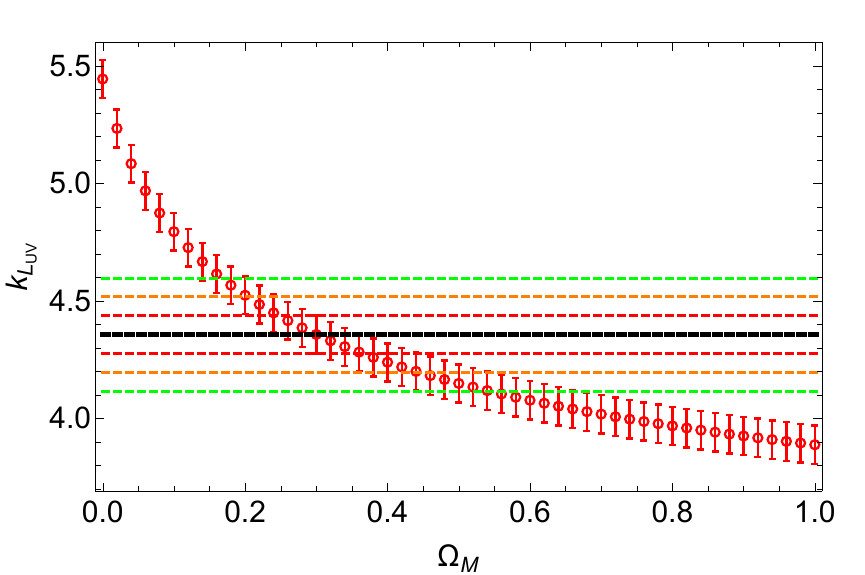}
    \caption{On the left panel, we present the behaviour of an evolutionary parameter obtained for the luminosity in the X-ray band $k_{L_{X}}$ as a function of $\Omega_M$. On the right panel, we present the results of an equivalent computation but for the luminosity in the UV band. The error bars correspond to the 1 $\sigma$ uncertainty level. The black, thick line corresponds to the value of correction obtained for $\Omega_M=0.3$. While dotted red, orange, and green lines correspond to the 1, 2, and 3 $\sigma$ uncertainty level on k obtained for $\Omega_M=0.3$.}
    \label{fig:k}
\end{figure}

This relationship enables the integration of evolutionary correction into the cosmological fitting without assuming any value of $\Omega_M$ a priori. We present the functions $k_{L_{X}}$ and $k_{L_{UV}}$ in Fig. \ref{fig:k}.

\bibliography{bibliografia_4}

\end{document}